\documentclass[preprint,showpacs,aps]{revtex4}
\usepackage{dcolumn}
\usepackage{amssymb}
\usepackage{amsmath}
\usepackage{graphicx}

\newcommand{\doublefig}{.5\textwidth}

\begin{document}

\title{Two-dimensional discrete solitons in rotating lattices}
\author{Jes\'{u}s Cuevas,$^{1}$ Boris A. Malomed,$^{2}$ and P. G. Kevrekidis$%
^{3}$}
\address{$^{1}$Grupo de F\'{\i}sica No Lineal, Departamento de F\'{\i}sica
Aplicada I, Escuela Universitaria Polit\'{e}cnica, C/ Virgen de
\'{A}frica,
7, 41011 Sevilla, Spain\\
$^{2}$Department of Physical Electronics, School of Electrical
Engineering,
Faculty of Engineering, Tel Aviv University, Tel Aviv 69978, Israel\\
$^{3}$Department of Mathematics and Statistics, University of
Massachusetts, Amherst, Massachusetts 01003-4515, USA}

\begin{abstract}
We introduce a two-dimensional (2D) discrete nonlinear Schr\"{o}dinger
(DNLS) equation with self-attractive cubic nonlinearity in a rotating
reference frame. The model applies to a Bose-Einstein condensate stirred by
a rotating strong optical lattice, or light propagation in a twisted bundle
of nonlinear fibers. Two species of localized states are constructed:
off-axis fundamental solitons (FSs), placed at distance $R$ from the
rotation pivot, and on-axis ($R=0$) vortex solitons (VSs), with vorticities $%
S=1$ and $2$. At a fixed value of rotation frequency $\Omega $, a stability
interval for the FSs is found in terms of the lattice coupling constant $C$, $%
0<C<C_{\mathrm{cr}}(R)$, with monotonically decreasing $C_{\mathrm{cr}}(R)$.
VSs with $S=1$ have a stability interval, $\tilde{C}_{\mathrm{cr}%
}^{(S=1)}(\Omega )<C<C_{\mathrm{cr}}^{(S=1)}(\Omega )$, which exists for $%
\Omega $ below a certain critical value, $\Omega _{\mathrm{cr}}^{(S=1)}$.
This implies that the VSs with $S=1$ are \emph{destabilized} in the
weak-coupling limit by the rotation. On the contrary, VSs with $S=2$, that
are known to be unstable in the standard DNLS equation, with $\Omega =0$,
are \emph{stabilized} by the rotation in region $0<C<C_{\mathrm{cr}}^{(S=2)}$%
, with $C_{\mathrm{cr}}^{(S=2)}$ growing as a function of $\Omega $.
Quadrupole and octupole on-axis solitons are considered too, their stability
regions being weakly affected by $\Omega \neq 0$.
\end{abstract}

 \maketitle

\section{Introduction}

Discrete dynamical systems represented by nonlinear lattices in
one, two, and three dimensions constitute a class of models which
are of fundamental interest by themselves, and, simultaneously,
they find applications of paramount importance in various fields
of physics. One such example is known in nonlinear optics, where
the one-dimensional (1D) discrete nonlinear Schr\"{o}dinger (DNLS)
equation was predicted by Christodoulides and Joseph to support
fundamental discrete solitons \cite{1988} with an even profile.
Such nonlinear structures
were later created experimentally in an array of
semiconductor
waveguides \cite{Yaron}. Subsequently, stable odd solitons, alias \textit{%
twisted localized modes}, were predicted and studied in detail in the same
model \cite{TLM}, as well as in an array of photorefractive waveguides with
photovoltaic nonlinearity \cite{TLMphotorefr-theory}.

Another experimental realization of dynamical lattices in the optical domain
is possible in photorefractive crystals, where a quasi-discrete setting can
be induced by counterpropagating laser beams illuminating the crystal in the
normal polarization, while the probe beam, which can sustain solitary waves,
is launched in the extraordinary polarization. The difference from the array
of waveguides fabricated in silica or in a semiconductor material is that
the photorefractive nonlinearity is saturable, rather than cubic. This
method of the creation of \textit{photonic lattices} was proposed in Ref.
\cite{photoinduction}, and results obtained by means of the technique were
reviewed in Refs. \cite{PhotorefrReviews}. In particular, both fundamental
and twisted solitons in the 1D lattice were reported in Ref. \cite%
{TLMphotorefr}, and fundamental solitons (FSs) in the 2D lattice were
created too \cite{Photorefr2Dsolitons}, as well as 2D vortex solitons (VSs)
in the same setting \cite{PhotorefrVortex}.

Recently, the progress in the technology of writing permanent arrays of
channels in silica slabs has made it possible to create 2D waveguiding
lattices with the cubic nonlinearity, which emulate a bundle of nonlinear
optical fibers with linear coupling in the transverse plane \cite%
{silica-bundle}. In particular, spatial lattice solitons of the surface and
corner types were reported in this setting \cite{silica-bundle-surface}
(surface solitons were reported too in the 2D photonic lattice in a
photorefractive crystal \cite{Zhigang-surface}). Thus, 1D and 2D discrete
dynamical models have a potential for further applications to nonlinear
optical media of various types.

Another natural realization for the lattice systems is provided by a
Bose-Einstein condensate (BEC) trapped in an optical lattice (OL). If the OL
is strong enough, the underlying Gross-Pitaevskii equation (GPE)\ for the
wave function in the continuum may be approximated by its DNLS counterpart
\cite{BEC,Wannier}; the relevance of the discrete model in this setting has
also been demonstrated experimentally in 1D by testing its predictions
experimentally, see, e.g., Ref. \cite{BECexperiment}. Higher-dimensional OLs
can be easily created too \cite{BECexperiment2}. Thus, the relevant DNLS
equation may be one-, two-, or three-dimensional; in particular, various
species of 3D discrete solitons, including those with intrinsic vorticity,
have been predicted in this setting \cite{3D}, and their stability has been
systematically analyzed \cite{pelin}. Still another implementation of the
DNLS lattice in the space of any dimension, from one to three, is possible
in terms of a crystal of microcavities trapping photons \cite{microcavity}
or polaritons \cite{polaritons}.

Discrete solitons of various kinds have been studied in detail theoretically
in 1D, 2D, and 3D versions of the DNLS equation, see an earlier review \cite%
{PanosReview} and the more recent works mentioned above. As mentioned above,
some of these solitons have been created experimentally in optical media
equipped with fabricated or photoinduced lattices. All these localized
states have their counterparts in continuum models with periodic potentials
that emulate the lattices. In particular, 2D solitons of both the
fundamental and vortex types, which are unstable in uniform continua with
the cubic self-focusing nonlinearity can be readily stabilized by the
periodic OL potential \cite{BBB}; for the stabilization of FSs, a quasi-1D
potential is sufficient, instead of its full 2D counterpart \cite{BBB2}.
Vortices are unstable too in the uniform space with the saturable
nonlinearity \cite{Dima}, in which case they can also be stabilized by the
periodic potential \cite{PhotorefrVortex}. The relevance of these
stabilization mechanisms in the continuum was also demonstrated for
higher-order vortex solitons, and so-called \textit{supervortices}, i.e.,
arrays of compact vortices with global vorticity imprinted onto the array,
under both cubic and saturable nonlinearities \cite{supervortex}.

Recently, it was shown that 2D solitons obeying the GPE in the 2D continuum
can also be supported by a \emph{rotating} OL \cite{HS,Barcelona}. These
solitons may be fully localized (spot-shaped) solutions to the equation with
the self-focusing/attractive cubic nonlinearity, placed at some distance
from the rotation pivot and revolving in sync with the holding 2D lattice.
In particular, the soliton can be placed at a local minimum of the rotating
potential, while the pivot is set at a local maximum. These \textit{%
co-rotating} strongly localized solitons are stable provided that the
rotation frequency, $\Omega $, does not exceed a critical value, $\left(
\Omega _{\mathrm{cr}}\right) _{\min }$. In the same model, but with a
rapidly rotating OL, stable \textit{ring-shaped} solitons (with zero
vorticity), i.e., objects localized along the radius but delocalized in the
azimuthal direction, have been found too, for $\Omega $ exceeding another
critical value, $\left( \Omega _{\mathrm{cr}}\right) _{\max }$. Note that
the model does not support any stable pattern in interval $\left( \Omega _{%
\mathrm{cr}}\right) _{\min }<\Omega <$ $\left( \Omega _{\mathrm{cr}}\right)
_{\max }$ \cite{HS}. On the other hand, stable ring-shaped states, with both
zero and nonzero vorticity, have been found in the model with the repulsive
cubic nonlinearity and rotating quasi-one-dimensional (periodic) potential,
if $\Omega $ exceeds a respective critical value. Obviously, the latter
model does not give rise to any localized state in the absence of the
rotation.

A rotating OL can be easily implemented in BEC\ experiments \cite{rotating};
then, if the OL is strong enough, it is natural to approximate the GPE in
the co-rotating reference frame by an appropriate variety of the 2D DNLS
equation. In addition to that, such a model may also describe the light
propagation in a \textit{twisted bundle} of nonlinear optical fibers,
linearly coupled in the transverse plane by tunneling of light between the
fiber cores. The objective of the present work is to introduce a model of
the rotating discrete lattice, and find stable discrete solitons in it, both
FSs and VSs.

The paper is organized as follows. In Sec. II, we formulate the model,
taking the underlying GPE in the reference frame co-rotating with the OL,
and replacing the continuum equation by its discrete version corresponding
to a strong periodic potential. Discrete FSs are considered in Sec. III. We
construct the solutions starting from the anti-continuum limit, which
corresponds to zero value of the coupling constant accounting for the linear
interaction between neighboring sites of the discrete lattice, $C=0$. A
family of FS solutions is constructed by continuation in $C$; their
stability is examined by computation of eigenfrequencies for infinitesimal
perturbations around the soliton, and verified by direct simulations of the
evolution of perturbed FSs. It is found that the FS, with its center located
at distance $R$ from the rotation pivot, is stable within an
interval $0<C<C_{%
\mathrm{cr}}(R)$, with $C_{\mathrm{cr}}$ decaying as a function of $R$.
Section III also includes a simple analytical approximation, which makes it
possible to explain the decrease of $C_{\mathrm{cr}}$ with the growth of $R$%
. In Sec. IV, we consider localized vortices (VSs), whose center
coincides with the rotation pivot. For the VS with vorticity $S=1$, the
stability region is found to be $\tilde{C}_{\mathrm{cr}}^{(S=1)}<C<C_{%
\mathrm{cr}}^{(S=1)}$, provided that the rotation frequency, $\Omega $, is
smaller than a critical value, $\Omega _{\mathrm{cr}}$ (the stability
interval shrinks to nil at $\Omega =\Omega _{\mathrm{cr}}$). Vortices with $%
S=2$ are considered too. While in the ordinary (nonrotating) DNLS model,
with $\Omega =0$, all VSs of the latter type are unstable \cite%
{higher-order-vortex}, we demonstrate that the rotation opens a \emph{%
stability window} for them, $0<C<C_{\mathrm{cr}}^{(S=2)}$, with $C_{\mathrm{%
cr}}^{(S=2)}$ growing as a function of $\Omega $. Direct
numerical simulations are also used to illustrate the
dynamical evolution of FSs and VSs when they are unstable.
Results obtained in
this work and related open problems are summarized in Sec. V.

\section{The model}

The starting point is the normalized 2D GPE,\ which includes the
potential in the form of an OL rotating at angular velocity
$\Omega $, and thus stirring a ``pancake"-shaped (quasi-flat)
Bose-Einstein condensate (BEC) trapped in a narrow gap between two
strongly repelling
optical sheets. Unlike the analysis performed in Refs. \cite{HS} and \cite%
{Barcelona}, where simulations were run in the laboratory reference frame,
here we write the GPE in the reference frame co-rotating with the lattice,
hence the potential does not contain explicit time dependence:
\begin{equation}
i\frac{\partial \psi }{\partial t}=-\left( \frac{1}{2}\nabla ^{2}+\Omega
\hat{L}_{z}\right) \psi -\epsilon \left[ \cos \left( k\left( x-\xi \right)
\right) +\cos (k\left( y-\upsilon \right) )\right] \psi +\sigma |\psi
|^{2}\psi .  \label{GPE}
\end{equation}%
Here, $\hat{L}_{z}=i(x\partial _{y}-y\partial _{x})\equiv i\partial _{\theta
}$ is the operator of the $z$-component of the orbital momentum ($\theta $
is the polar angle), and $\sigma $ determines the sign of the interaction,
attractive ($\sigma =-1$) or repulsive ($\sigma =+1$). Constants $\xi $ and $%
\upsilon $ determine a possible shift of the lattice with respect to the
rotation pivot. Note that Eq. (\ref{GPE}) does not contain any additional
trapping potential, as we are interested in solutions localized under the
action of the OL.

It is more convenient to shift the origin of the Cartesian coordinates to a
lattice node, thus replacing Eq. (\ref{GPE}) by the translated form:
\begin{equation}
i\frac{\partial \psi }{\partial t}=-\left[ \frac{1}{2}\nabla ^{2}+i\Omega
(\left( x+\xi \right) \partial _{y}-\left( y+\upsilon \right) \partial _{x})%
\right] \psi -\epsilon \left[ \cos \left( kx\right) +\cos (ky)\right] \psi
+\sigma |\psi |^{2}\psi .  \label{shifted}
\end{equation}%
As shown in a general form in Ref. \cite{Wannier}, a discrete model, which
corresponds to the limit of a very deep OL, can be derived from the
underlying GPE in the \textit{tight-binding approximation}. Eventually, it
amounts to a
straightforward discretization of the GPE. Thus, the discrete
counterpart of Eq. (\ref{shifted}) is
\begin{eqnarray}
i\frac{d\psi _{m,n}}{dt} &=&-\frac{C}{2}\left\{ \left( \psi _{m+1,n}+\psi
_{m-1,n}+\psi _{m,n+1}+\psi _{m,n-1}-4\psi _{m,n}\right) \right.  \notag \\
&&\left. -i\Omega \left[ \left( m+\xi \right) \left( \psi _{m,n+1}-\psi
_{m,n-1}\right) -\left( n+\upsilon \right) \left( \psi _{m+1,n}-\psi
_{m-1,n}\right) \right] \right\} +\sigma |\psi _{m,n}|^{2}\psi _{m,n}~,
\label{discrete}
\end{eqnarray}%
where $\left( m,n\right) $ are discrete coordinates, and $C>0$ is the
corresponding coupling constant, which accounts for the linear tunneling of
atoms between BEC\ droplets trapped in deep nodes of the lattice. As
mentioned above, Eq. (\ref{discrete}) may also describe a twisted bundle of
nonlinear optical fibers linearly coupled by the tunneling of light in the
transverse plane, $\left( m,n\right) $. In that case, $t$ is the propagation
distance along the fiber, and only $\sigma =-1$, i.e., the
attractive/focusing nonlinearity) is the relevant choice.

Equation (\ref{discrete}) conserves the norm and Hamiltonian,
\begin{equation}
N=\sum_{m,n}\left\vert \psi _{m,n}\right\vert ^{2},  \label{N}
\end{equation}%
\begin{eqnarray}
H &=&\sum_{m,n}\left\{ \frac{C}{2}\left( \left\vert \psi _{m+1,n}-\psi
_{m,n}\right\vert ^{2}+\left\vert \psi _{m,n+1}-\psi _{m,n}\right\vert
^{2}\right) +\frac{1}{2}\sigma \left\vert \psi _{m,n}\right\vert ^{4}\right.
\notag \\
&&-\frac{iC}{4}\Omega \left[ \left( m+\xi \right) \left( \psi _{m,n}^{\ast
}\left( \psi _{m,n+1}-\psi _{m,n-1}\right) -\psi _{m,n}\left( \psi
_{m,n+1}^{\ast }-\psi _{m,n-1}^{\ast }\right) \right) \right.  \notag \\
&&\left. \left. -\left( n+\upsilon \right) \left( \psi _{m,n}^{\ast }\left(
\psi _{m+1,n}-\psi _{m-1,n}\right) -\psi _{m,n}\left( \psi _{m+1,n}^{\ast
}-\psi _{m-1,n}^{\ast }\right) \right) \right] \right\} .  \label{H}
\end{eqnarray}%
In addition to $C$, the discrete model contains three irreducible
parameters: $\Omega $, which takes values $0<\Omega <\infty $, and the
coordinates of the pivot displacement, $\left( \xi ,\upsilon \right) $,
which take values $0\leq \xi ,\upsilon <1$, plus the sign parameter, $\sigma
=\pm 1$. As in the usual 2D DNLS equation (with $\Omega =0$), values $\sigma
=\pm 1$ in Eq. (\ref{discrete}) may be transformed into each other by the
\textit{staggering transformation}, $\psi _{m.n}\rightarrow (-1)^{m+n}\psi
_{m,n}$, therefore we fix $\sigma \equiv -1$ (self-attraction).

Our first objective is to find stationary localized solutions to Eq. (\ref%
{discrete}) in the form of FSs (fundamental solitons) and VSs\ (vortex
solitons). To this end, we substitute the standing wave ansatz, $\psi
_{m,n}=e^{i\Lambda t}\phi _{m,n}$, where $-\Lambda $ is the normalized
chemical potential, in terms of the underlying BEC model; then, the
stationary lattice field $\phi _{m,n}$ obeys the equation
\begin{eqnarray}
\Lambda \phi _{m,n} &=&\frac{C}{2}\left( \phi _{m+1,n}+\phi _{m-1,n}+\phi
_{m,n+1}+\phi _{m,n-1}-4\phi _{m,n}\right)  \notag \\
&&+i\frac{C}{2}\Omega \left[ \left( m+\xi \right) \left( \phi _{m,n+1}-\phi
_{m,n-1}\right) -\left( n+\upsilon \right) \left( \phi _{m+1,n}-\phi
_{m-1,n}\right) \right] +|\phi _{m,n}|^{2}\phi _{m,n}~.  \label{phi}
\end{eqnarray}%
Note that solutions for $\phi _{m,n}$ are complex, unless $\Omega =0$. The
second objective will be to examine the stability of the discrete solitons,
assuming small perturbations in the form of $\delta \psi _{m,n}\sim \exp
\left( i\Lambda t+i\lambda t\right) $, the onset of instability indicated by
the emergence of $\mathrm{Im}(\lambda )\neq 0$. The evolution of unstable
solitons will be examined by means of direct simulations of Eq. (\ref%
{discrete}).

To parameterize the soliton families, we fix the scales by setting $\Lambda
\equiv 1$; obviously, with $C>0$, only positive $\Lambda $ may give rise to
localized solutions), while $C$ will be varied.

\section{Fundamental solitons}

The rotation makes the discrete lattice inhomogeneous, hence properties of
solitons strongly depend of the location of their centers. Without the
rotation, Eq. (\ref{discrete}) amounts to the ordinary 2D DNLS\ equation, in
which various species of discrete solitons and their stability have been
studied in detail. In particular, FSs, which are represented by real
solutions, are stable at $C\leq C_{\mathrm{cr}}=2\Lambda \equiv 2$ \cite%
{PanosReview}. The onset of their instability is accounted for by a pair of
eigenfrequencies of small perturbations with finite imaginary and zero real
parts, i.e., the instability leads to the exponential growth of
perturbations. Accordingly, numerical simulations of the instability
development demonstrate spontaneous transformation of unstable FSs into
lattice breathers \cite{gfb06}. In this section, we first report numerical
results obtained for the stability of FSs in the model with $\Omega \neq 0$,
and then present an analytical estimate that may explain numerical findings.

\subsection{Numerical results}

Our analysis aimed to determine the stability border for the FSs, $C_{%
\mathrm{cr}}$, for each set of values of the discrete coordinates of the
soliton's center, $\{m_{0},n_{0}\}$. Here we present results for angular
velocity $\Omega =0.1$ and zero pivot displacement, $\xi =\upsilon =0$ [%
in Ref. \cite{HS}, the rotation pivot was fixed at a local maximum
of the potential in Eq. (\ref{shifted}), which corresponds to
setting $\xi =\upsilon =1/2$ in Eqs. (\ref{discrete}) and
(\ref{phi}), and the center of the soliton trapped in the lattice
was placed at a local minimum closest to the pivot, which would mean
$\{m_{0},n_{0}\}=\{0,0\}$]. This choice makes it possible to explore
the existence and stability of FSs in a clear form, while larger
values of $\Omega $ give rise to a resonance with linear
lattice modes, leading to Wannier-Stark ladders and hybrid solitons \cite%
{hybrid} and making the continuation in $C$ and identification of $C_{%
\mathrm{cr}}$ difficult. We carried out the calculations on the lattice of
size $21\times 21$, since already for this case, the lattice was for all the
considered cases sufficiently wider than the very localized FS structures of
interest. To avoid effects of the boundaries, the range of the
soliton-center's coordinates was restricted to $\left\vert m_{0}\right\vert
,\left\vert n_{0}\right\vert \leq 8$.

The FS solutions were looked for starting at point $C=0$, i.e., at the
so-called anti-continuum limit \cite{PanosReview}. In this limit, the FS is
seeded by using a nonzero value of the
field at a single point, the center
of the FS, $\phi _{m,n}^{(C=0)}=\delta _{m,m_{0}}\delta _{n,n_{0}}$%
; obviously, this expression satisfies Eq. (\ref{phi}) with $\Lambda =1$ and
$C=0$. After the branch of the FS solutions had been found by the
continuation in $C$, its stability was quantified through the computation of
eigenfrequencies
of small perturbations, using the equation linearized about the
FS.

\begin{figure}[tbp]
\begin{center}
\begin{tabular}{cc}
\includegraphics[width=\doublefig]{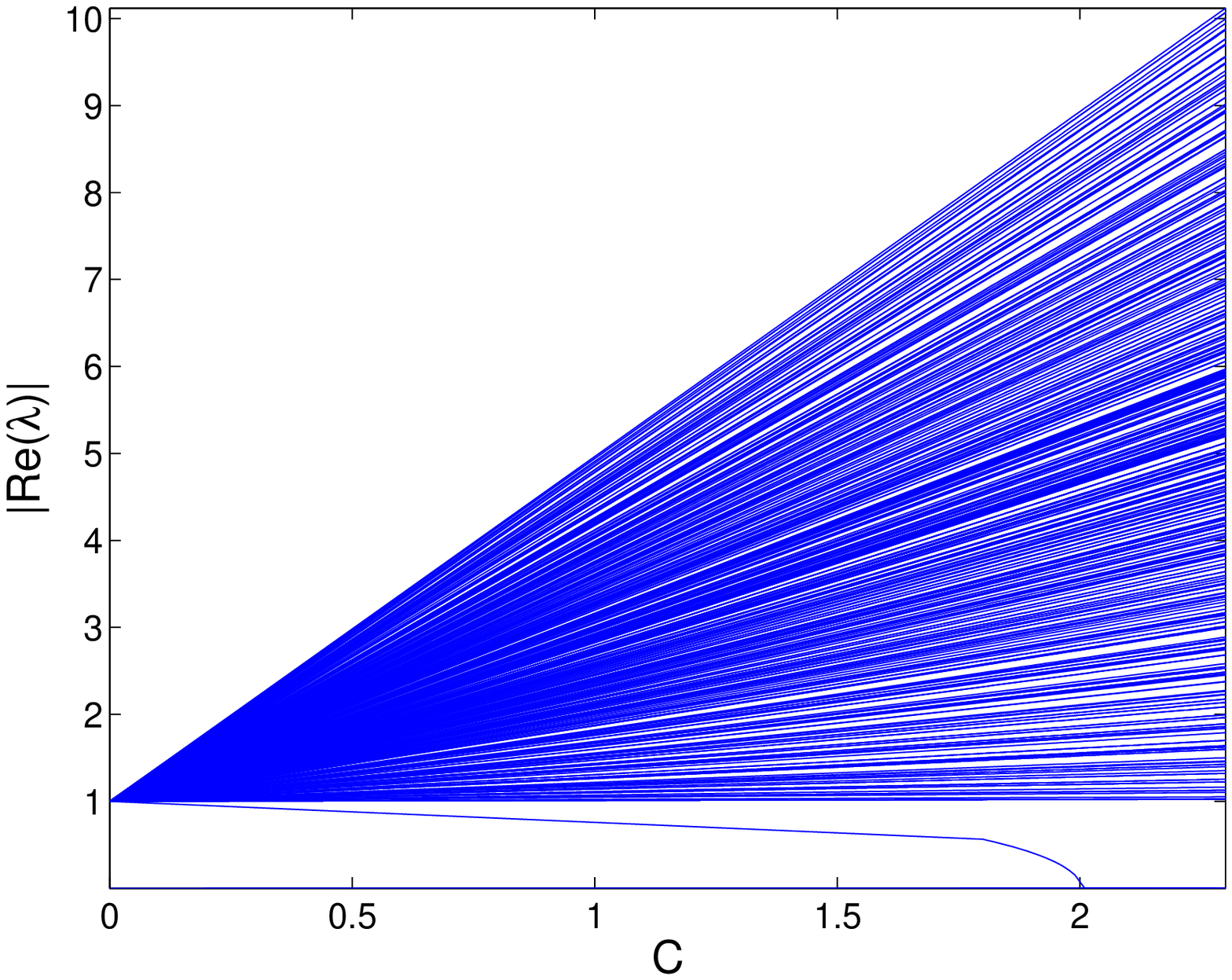} & \includegraphics[width=%
\doublefig]{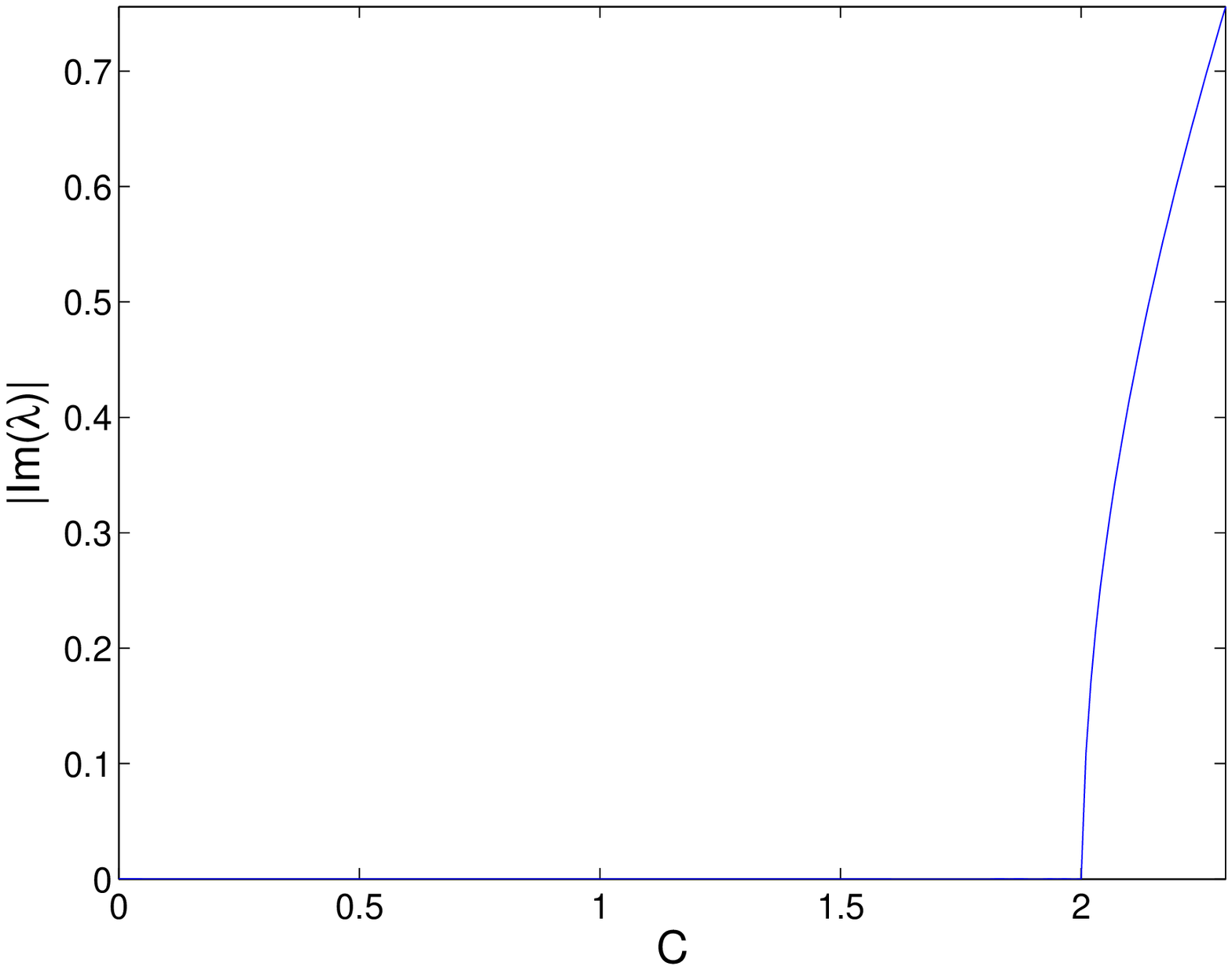} \\
\includegraphics[width=\doublefig]{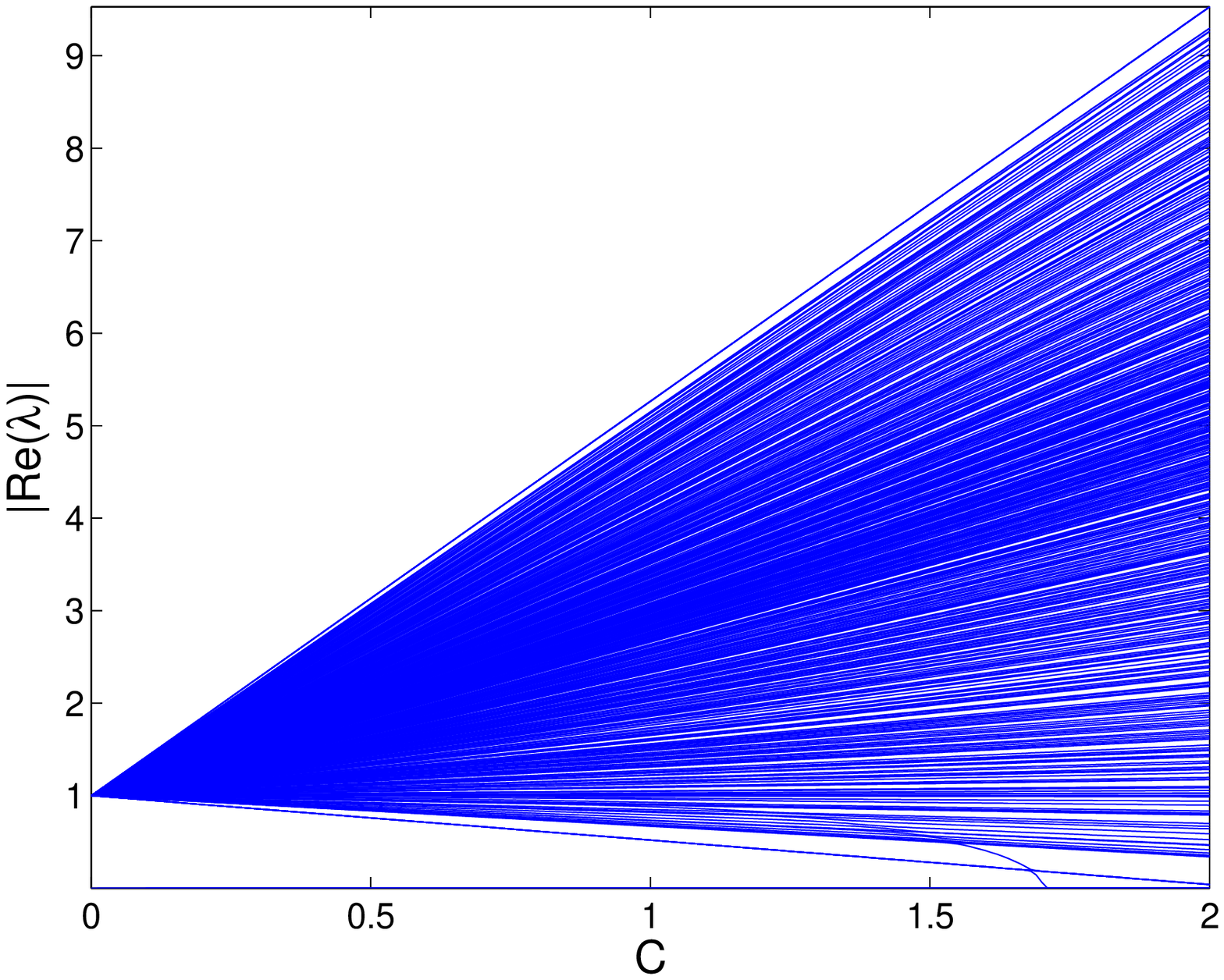} & \includegraphics[width=%
\doublefig]{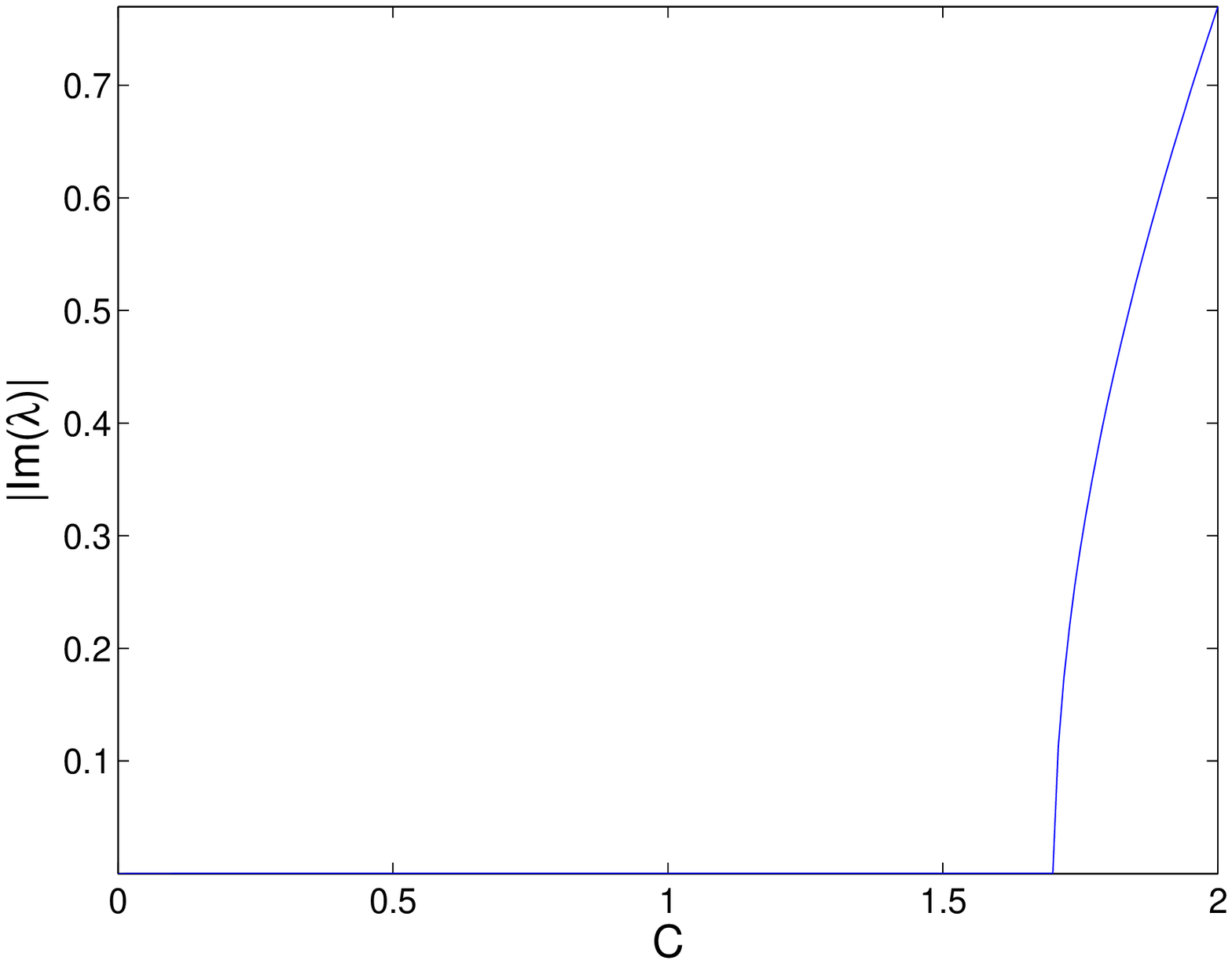}%
\end{tabular}%
\end{center}
\caption{(Color online) Plots of real and imaginary parts (left and
right panels) of the eigenfrequencies of linearization around fundamental
solitons in the ordinary non-rotating lattice, with $\Omega =0$ (top
panels), and for their counterparts in the present model (with $\Omega =0.1$%
). In the latter case, the soliton is centered at $(m_{0}=3,n_{0}=2)$.}
\label{fig:stabfsol}
\end{figure}

\begin{figure}[tbp]
\begin{center}
\begin{tabular}{cc}
\includegraphics[width=\doublefig]{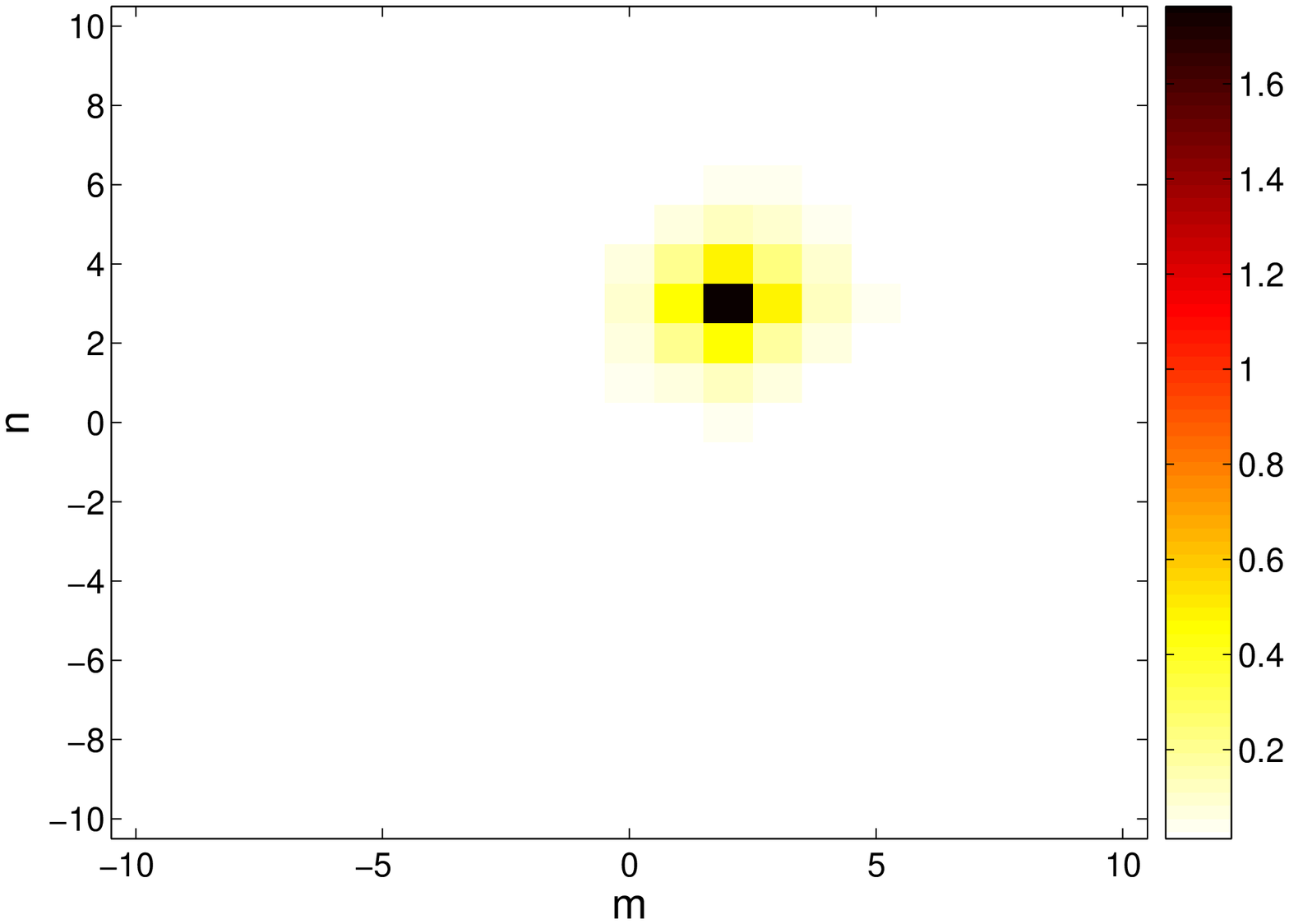} & \includegraphics[width=%
\doublefig]{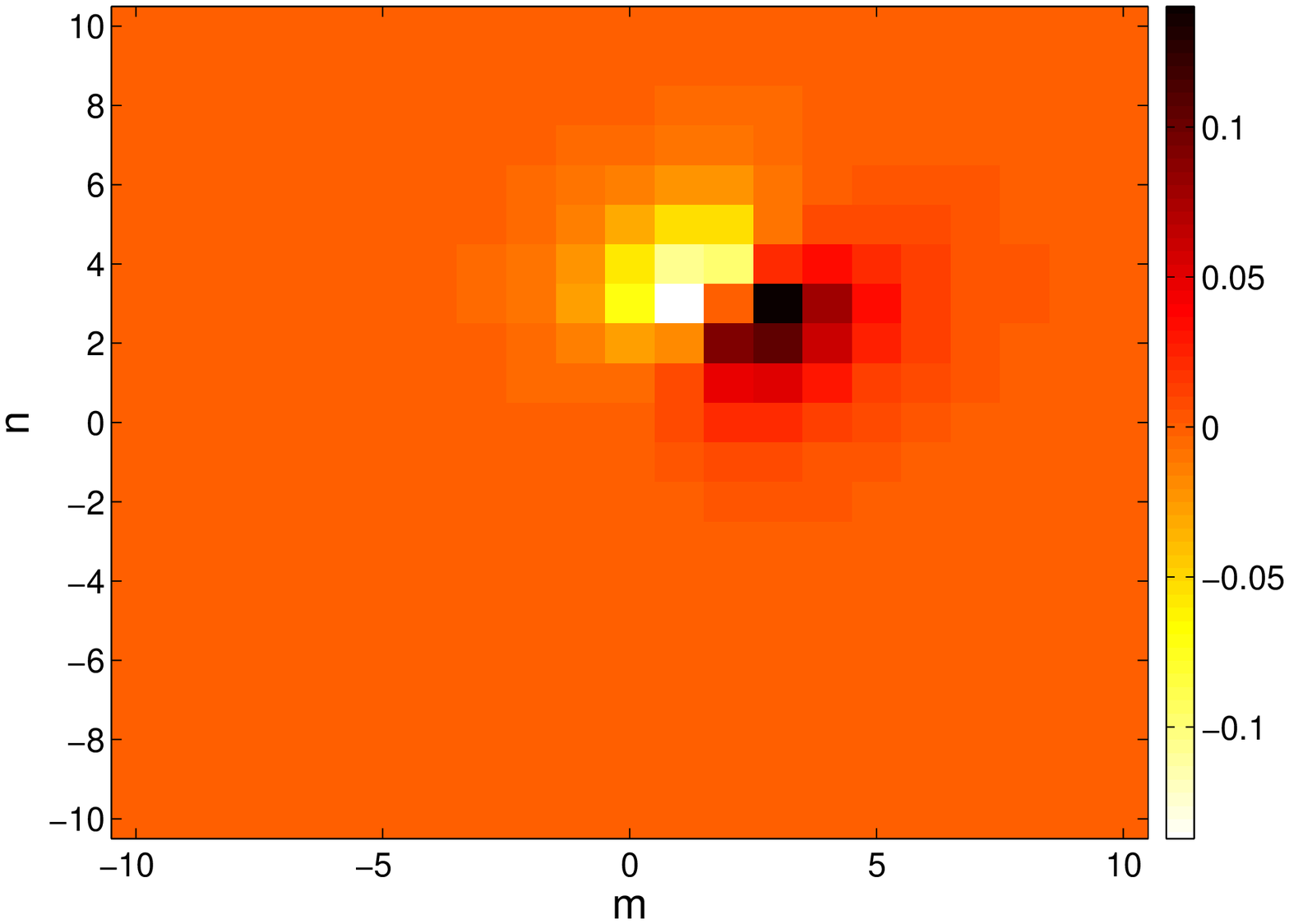} \\
\multicolumn{2}{c}{\includegraphics[width=\doublefig]{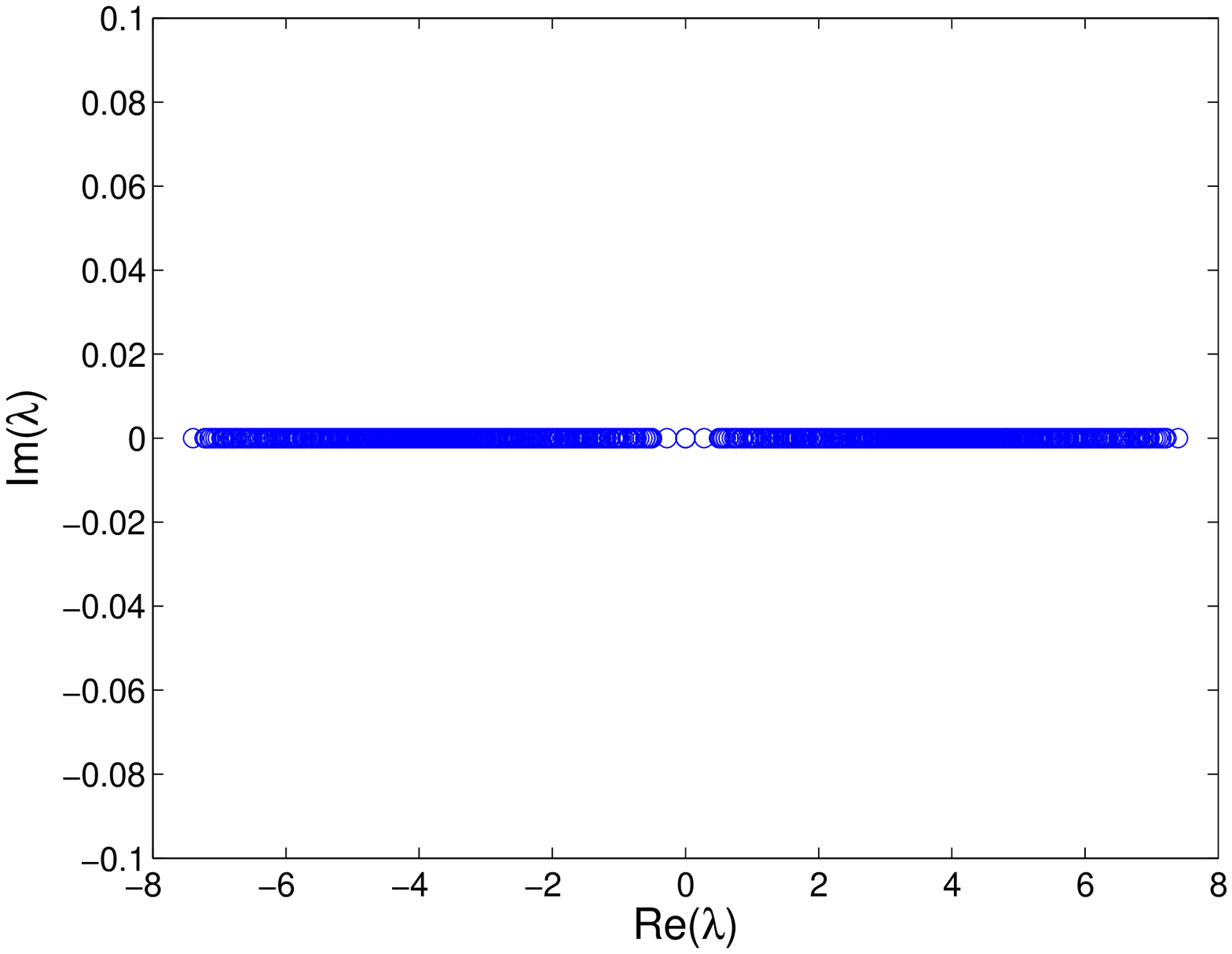}}%
\end{tabular}%
\end{center}
\caption{(Color online)\ Top panels: Contour plots of the real (left) and
imaginary (right) parts of a stable fundamental soliton centered at $m_{0}=3$%
, $n_{0}=2$, for $C=1.5$. Bottom panel: the complex plane of the stability
eigenvalues for this soliton.}
\label{fig:fsol1}
\end{figure}

\begin{figure}[tbp]
\begin{center}
\begin{tabular}{cc}
\includegraphics[width=\doublefig]{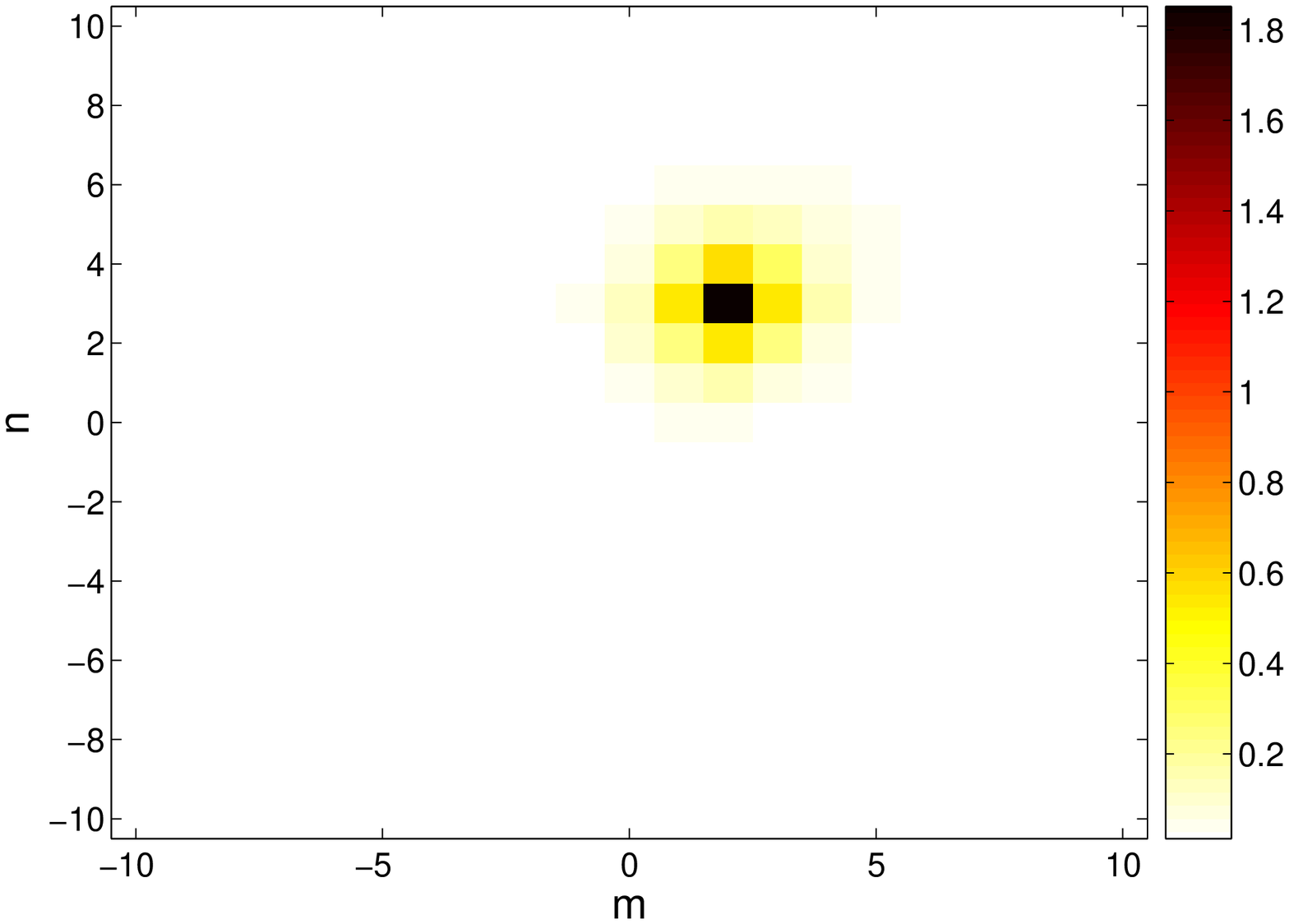} & \includegraphics[width=%
\doublefig]{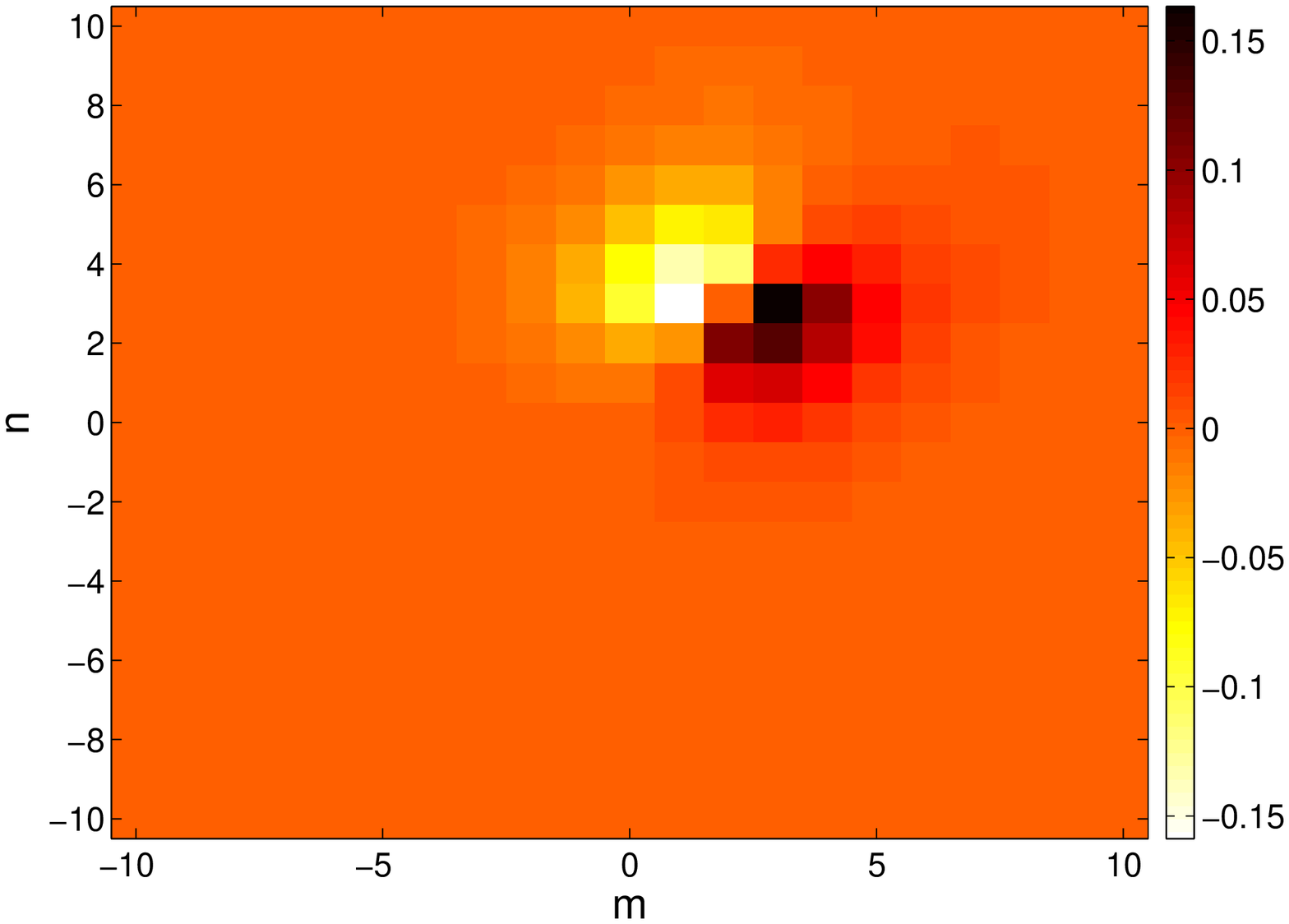} \\
\multicolumn{2}{c}{\includegraphics[width=\doublefig]{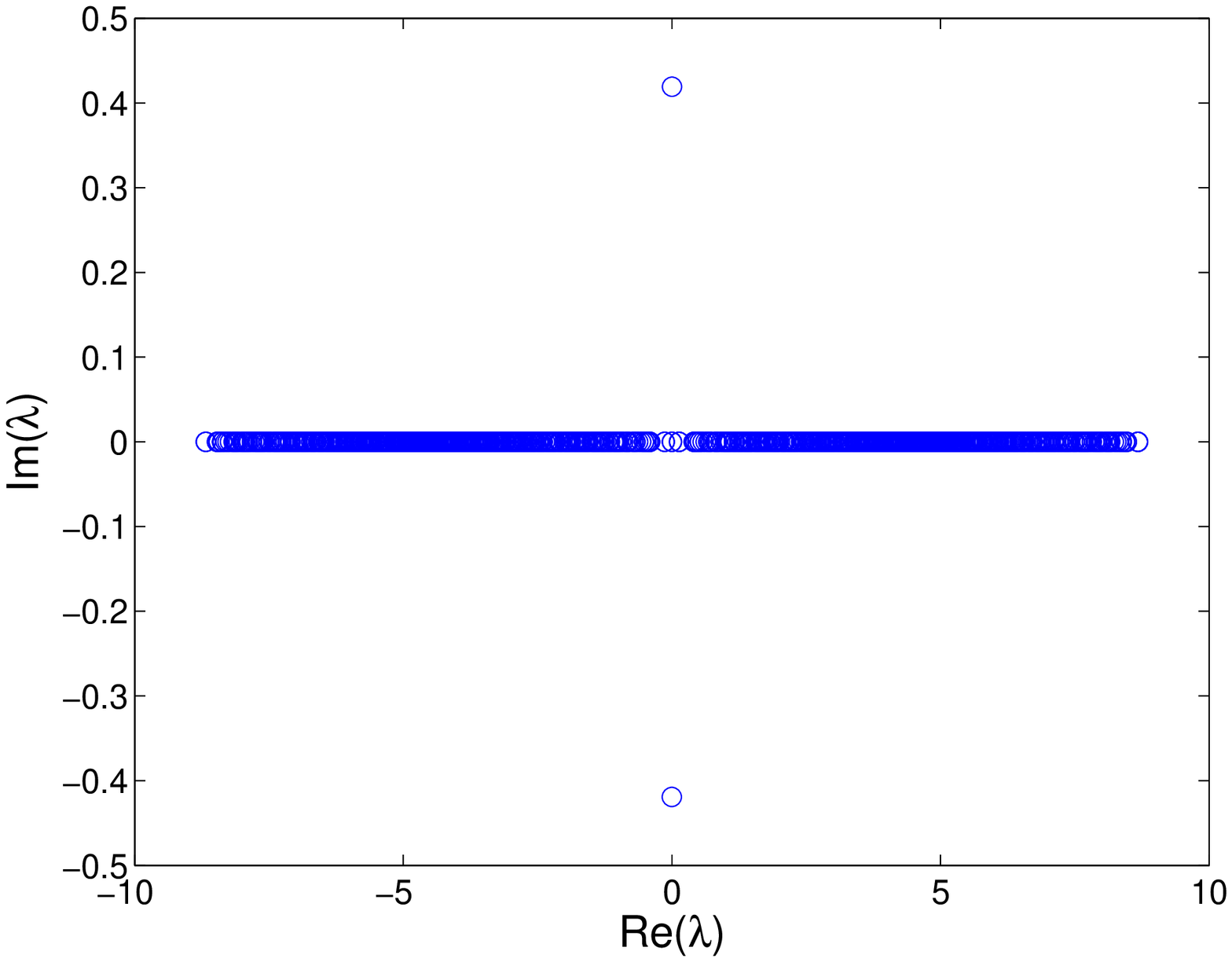}}%
\end{tabular}%
\end{center}
\caption{(Color online) The same as in Fig. \protect\ref{fig:fsol1}, but for
an unstable fundamental soliton, placed at the same position as in Fig.
\protect\ref{fig:fsol1}, at $C=1.8$.}
\label{fig:fsol2}
\end{figure}

Figure \ref{fig:stabfsol} displays a typical example of the thus found
dependence of eigenfrequencies $\lambda $ of small perturbations on the
lattice coupling constant. As said above, the instability corresponds to $%
\mathrm{Im}(\lambda )\neq 0$, for the FS with its center set at point ($%
m_{0}=3,n_{0}=2$); for comparison, the dependence of the instability growth
rate, i.e., $\left\vert \mathrm{Im}(\lambda )\right\vert $, on $C$ is also
shown for the FS in the usual DNLS model ($\Omega =0$) in the top panel of
the figure. It is seen that the instability sets in at $C=C_{\mathrm{cr}%
}=1.70$, which is smaller than the critical value in the ordinary model, $C_{%
\mathrm{cr}}^{(\Omega =0)}=2$. Typical examples of stable and unstable FSs
belonging to the family presented in Fig. \ref{fig:stabfsol} are displayed
in Figs. \ref{fig:fsol1} and \ref{fig:fsol2}.

\begin{figure}[tbp]
\begin{center}
\begin{tabular}{cc}
\includegraphics[width=\doublefig]{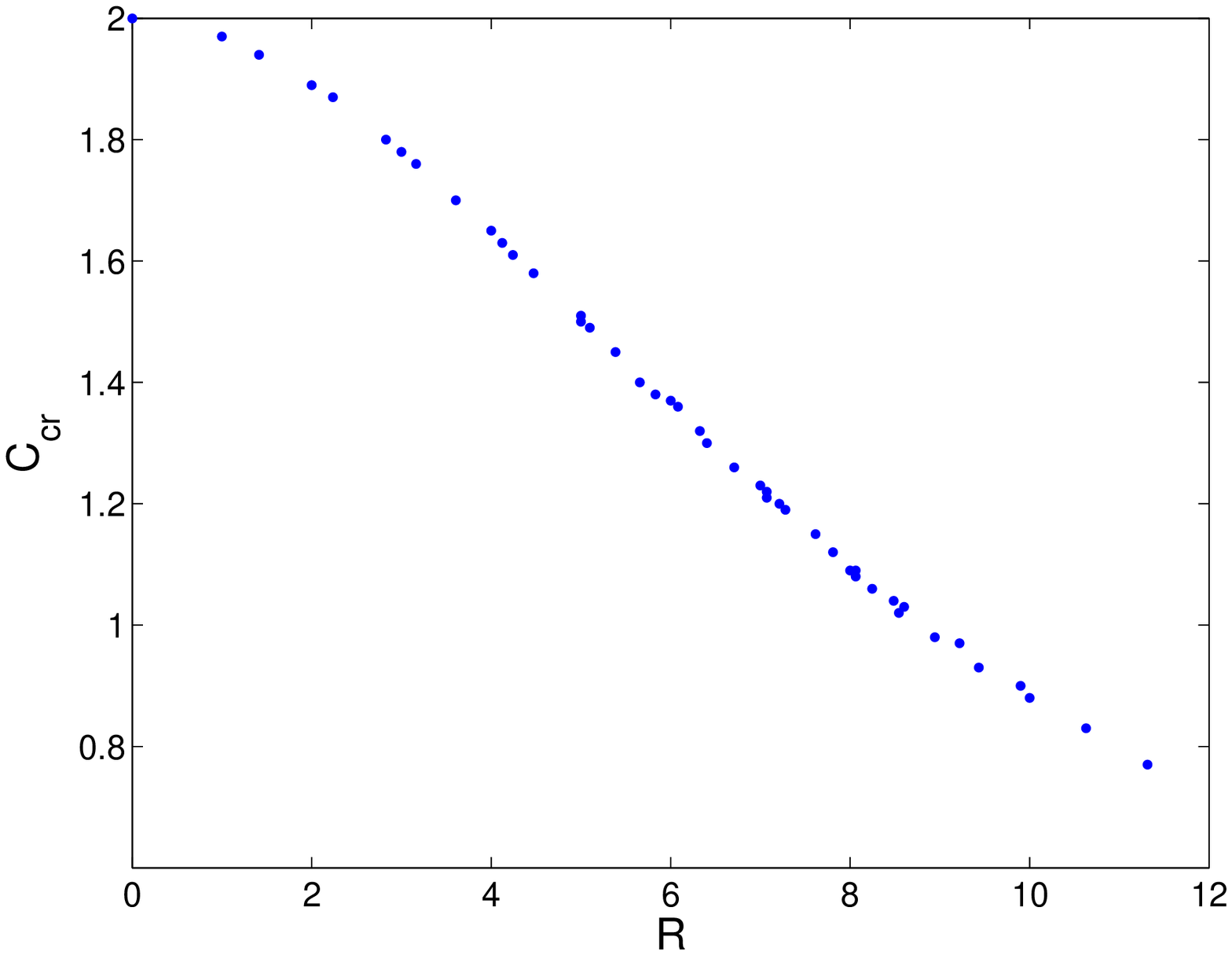} & \includegraphics[width=%
\doublefig]{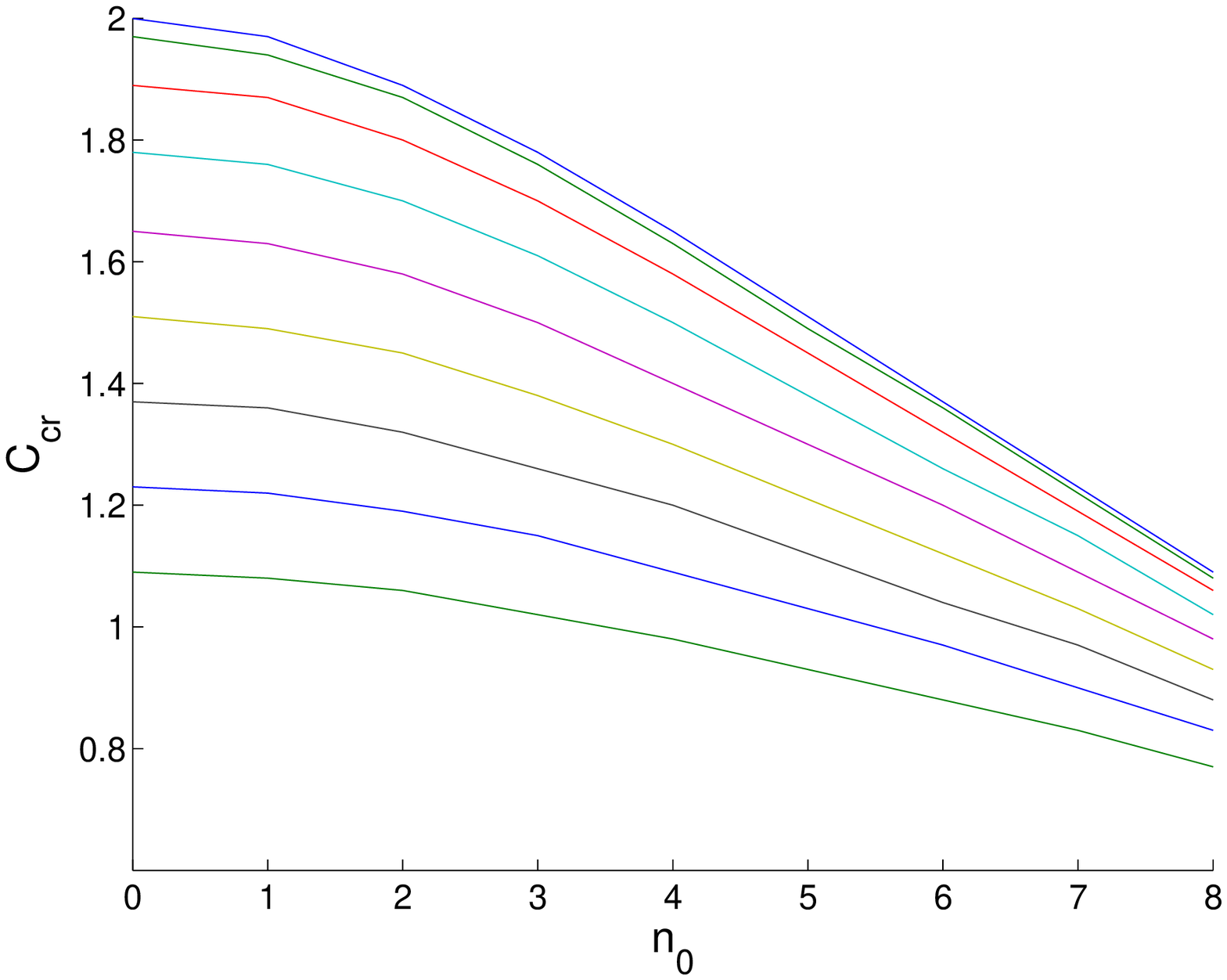}%
\end{tabular}%
\end{center}
\caption{(Color online) Dependence of the stability border for the
fundamental solitons, $C_{\mathrm{cr}}$, on the distance of the soliton's
center from the rotation pivot, $R$ (left panel), and on one of the center's
coordinates, $n_{0}$, for fixed $m_{0}$ (right panel). In the latter figure,
the value of $n_{0}\in \lbrack 0,8]$ increases progressively from top to
bottom.}
\label{fig:Ccr}
\end{figure}

Results obtained for the FSs placed at different positions are summarized in
Fig. \ref{fig:Ccr}, in the form of dependences of $C_{\mathrm{cr}}$ on the
distance of the FS's center from the pivot, $R\equiv \sqrt{%
m_{0}^{2}+n_{0}^{2}}$, and on one coordinate, $n_{0}$, while $m_{0}$ is
fixed. It is observed that $C_{\mathrm{cr}}$ \emph{monotonously decreases}
with $R$, starting from $C_{\mathrm{cr}}=2$ at $R=0$ (we recall again that $%
C_{\mathrm{cr}}^{(\Omega =0)}=2$ for the FS in the ordinary DNLS equation).
Note that there are different pairs $(m_{0},n_{0})$ which have equal values
of $R=\sqrt{m_{0}^{2}+n_{0}^{2}}$ and give slightly
different $C_{\mathrm{cr}}$. For
instance, for the pair $(5,0)$, $C_{\mathrm{cr}}=1.51$, while $C_{\mathrm{cr}%
}=1.50$ for $(4,3)$. Hence, the stability depends on the two-dimensional
structure of the solution.

\begin{figure}[tbp]
\begin{center}
\includegraphics[width=\doublefig]{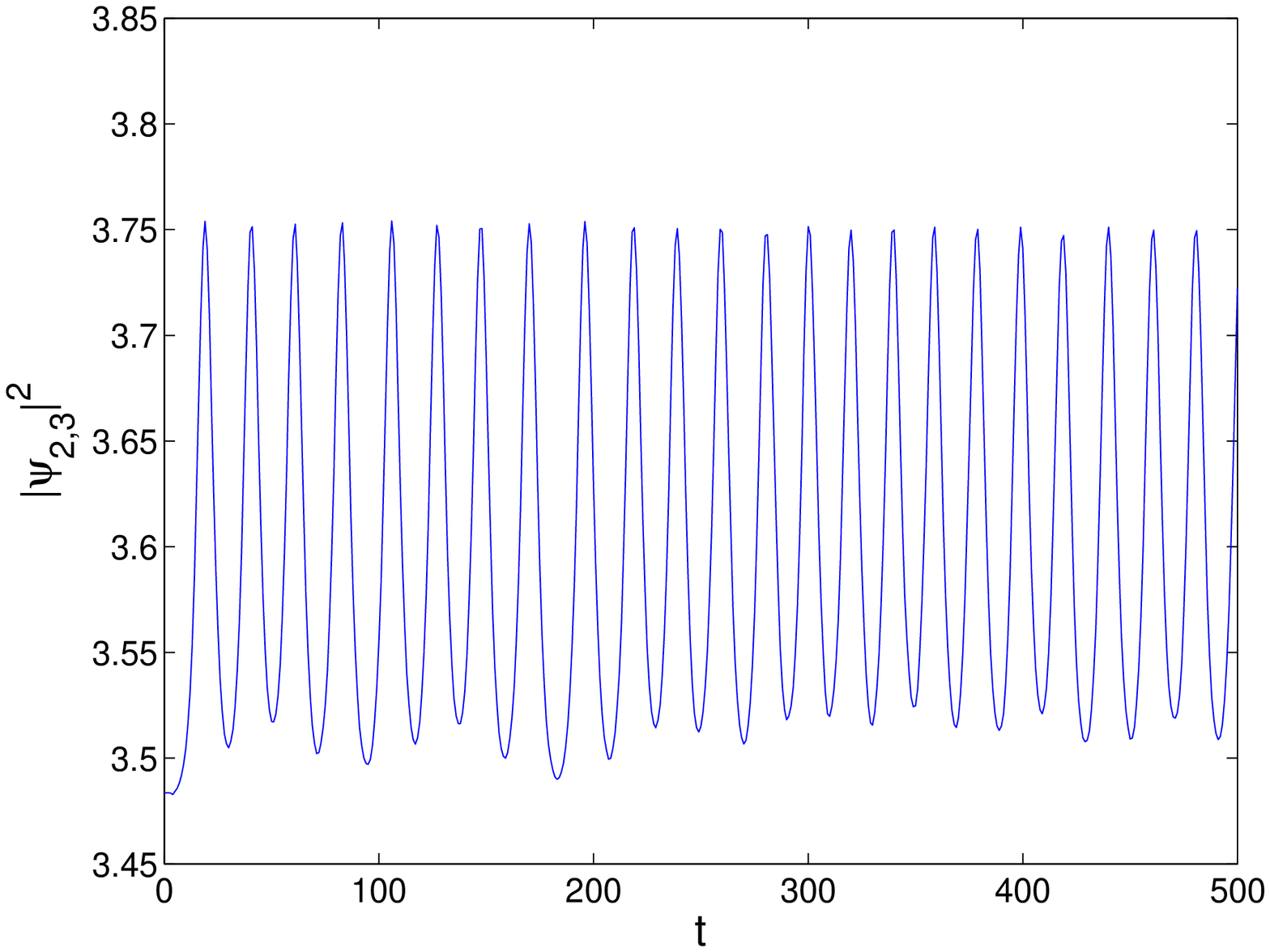}
\end{center}
\caption{(Color online) The time dependence of the squared amplitude of a
perturbed unstable soliton from Fig.~\protect\ref{fig:fsol2} at its center, $%
\left( m_{0}=3,n_{0}=2\right) $. Notice the robust oscillatory behavior
indicating the breathing nature of the resulting solution.}
\label{fig:pulsonfsol}
\end{figure}

Figures \ref{fig:stabfsol} and \ref{fig:fsol2} show that the FS is
destabilized, with the increase of $C$, through the appearance of a pair of
imaginary eigenfrequencies. Direct simulations of the dynamical evolution of
unstable FSs in the framework of Eq. (\ref{discrete}) demonstrate that the
instability does not destroy the solitary wave. Instead, it transforms the
waveform into a persistent breathing structure, see a typical example in
Fig. \ref{fig:pulsonfsol}.

\subsection{Analytical estimates}

The decrease of $C_{\mathrm{cr}}$ with the increase of the distance of the
FS from the pivot, $R$, which is the main feature revealed by the above
numerical analysis, as shown in Fig. \ref{fig:Ccr}, can be explained using
an estimate based on the quasi-continuum approximation. To this end, we note
that stationary solutions to the underlying continuum equation (\ref{GPE})
are looked for as $\psi =e^{i\Lambda t}\phi (x,y)$, with the function $\phi $
obeying the stationary equation,%
\begin{equation}
\Lambda \phi =-C\left( \frac{1}{2}\nabla ^{2}+\Omega \hat{L}_{z}\right) \phi
-\epsilon \left[ \cos \left( kx\right) +\cos (ky)\right] \Phi -\Phi ^{3}.
\label{Phi}
\end{equation}%
Here, following the discrete model, we have set $\xi =\upsilon =0$, $\sigma
=-1$, and explicitly introduced the spatial-scale parameter, $1/\sqrt{C}$,
which is a counterpart of the lattice coupling constant in Eq. (\ref%
{discrete}). Next, we assume the presence of a soliton with amplitude $A$
and intrinsic size $l$, whose center is located at distance $R$ from the
rotation pivot (tantamount to the origin, in the present case). First,
demanding a balance between terms $\Lambda \phi $, $\nabla ^{2}\phi ,$ and $%
\phi ^{3}$ in Eq. (\ref{Phi}), and estimating them, respectively, as $%
\Lambda A$, $A/l^{2}$, and $A^{3}$, we conclude, in the lowest
approximation, that%
\begin{equation}
\left\vert \Lambda \right\vert \sim Cl^{-2}\sim A^{2}.  \label{est1}
\end{equation}

Further, the soliton as a whole will be in equilibrium relative to the
rotating lattice potential if the action of the centrifugal force, generated
by term $\sim \Omega $ in Eq. (\ref{Phi}), is compensated by the force of
pinning to the periodic potential. The estimate of the latter condition
yields $C\Omega R/l\sim \epsilon $. Substituting here $l\sim \sqrt{%
C/\left\vert \Lambda \right\vert }$, as per Eq. (\ref{est1}), we arrive at a
final estimate, $R\sim \left( \epsilon /\sqrt{\left\vert \Lambda \right\vert
}\Omega \right) C^{-1/2}$, which predicts dependence $C_{\mathrm{cr}}\sim
1/R^{2}$ ($C_{\mathrm{cr}}$ is realized here as the largest value of $C$
that can provide for the balance between the centrifugal and pinning forces
at given $R$). The latter dependence is qualitatively consistent with the
numerical findings showing the decrease of $C_{\mathrm{cr}}$ with $R$ in
Fig. \ref{fig:Ccr} (at very large $C$, when the analytical formula predicts
small $R$, it is irrelevant, as the above consideration tacitly assumed that
the size of the soliton was essentially smaller than the distance to the \
pivot, $l\ll R$; it is irrelevant too at very small $C$, as the
quasi-continuum approximation cannot be used in that essentially discrete
case).

Using the quasi-continuum approximation, it is also possible to estimate the
order of magnitude of the rotation frequency, in physical units. In the
application to the BEC, assuming the lattice spacing $\sim 1$ $\mu $m and
the condensate of $^{7}$Li or $^{85}$Rb, which provide for the possibility
of the attraction between atoms and, thus, the formation of solitons \cite%
{Li,Rb}, undoing of rescalings which cast the GPE in the normalized form of
Eq. (\ref{GPE}), we conclude that $\Omega =0.1$ corresponds, in physical
units, to the rotation frequency $\sim 100$ Hz or $10$ Hz, for lithium and
rubidium, respectively. As concerns the above-mentioned realization of the
model in terms of the twisted bundle of optical fibers, an estimate shows
that, for the carrier wavelength $\sim 1$ $\mu $m and separation between the
fibers in the bundle $\sim 10$ $\mu $m, $\Omega =0.1$ corresponds to the
twist pitch, which we define as a length at which the twist attains the
angle of $2\pi $, of the order of $5$ cm.

\section{Vortex solitons}

Discrete vortex solitons (VSs) of the 2D DNLS equation were
systematically developed in Ref. \cite{vort1} as complex
stationary solutions which feature a phase circulation of $2\pi S$
around the central point, at which the amplitude vanishes, with
integer $S$ identified as the vorticity. Prior to that,
time-periodic multibreather states that may feature a vortical
structure were found in 2D Hamiltonian lattice
dynamical models \cite{Aubry}. The center of the VS may coincide
with a site of the lattice, or may be located in the middle of a
lattice cell; the corresponding vortices are called ``crosses"
(alias rhombuses) and ``squares", respectively. The stability of
these states has been studied in detail, both for $S=1$
\cite{vort1,vort2} and $S>1$ \cite{higher-order-vortex,vort2}. In
particular, the VS crosses with $S=1$ (and, as above, with
$\Lambda $ fixed to be $1$) are stable in
the corresponding interval of values of the coupling constant, $C<C_{\mathrm{%
cr}}^{(S=1)}=0.781$, and the instability above this points transforms the VS
into an ordinary fundamental soliton, with $S=0$. While all VSs with $S=2$
are unstable in the ordinary 2D DNLS\ equation, the vortex solutions with $%
S=3$ have their stability interval (more narrow than for $S=1$), $C<C_{%
\mathrm{cr}}^{(S=3)}=0.198$. In all cases, the instability sets in via the
Hamiltonian Hopf bifurcation, represented by quartets of eigenvalues \cite%
{HH}.

We have constructed localized vortices (of the cross/rhombus type), with $S=1
$ and $S=2$, in the rotating-lattice model based on Eq. (\ref{discrete}),
and examined their stability. Unlike the above analysis of the FSs, we
consider here only \textit{on-axis} VSs, whose centers coincide with the
rotation pivot ($R=0$). We note that, while the FS with $R=0$ is virtually
identical to its counterpart in the ordinary model, with $\Omega =0$, for
VSs the situation is quite different; in particular, the VSs centered at the
rotation pivot feature quite nontrivial stability properties with the
increase of rotation frequency $\Omega $, see below. Therefore, while the
results for the FSs were displayed for $\Omega =0.1$ and growing $R$, in
this section we focus on $R=0$ but vary $\Omega $.

\subsection{Vortex solitons with $S=1$}

The first noteworthy effect of the rotation on the VS with $S=1$ is that,
for given $\Omega $, its stability region features not only the upper bound,
$C_{\mathrm{cr}}^{(S=1)}$, as in the usual DNLS model, but also a \emph{lower%
} one, $\tilde{C}_{\mathrm{cr}}^{(S=1)}$:
\begin{equation}
\tilde{C}_{\mathrm{cr}}^{(S=1)}<C<C_{\mathrm{cr}}^{(S=1)}.
\label{VSstability}
\end{equation}

\begin{figure}[tbp]
\begin{center}
\begin{tabular}{cc}
\includegraphics[width=\doublefig]{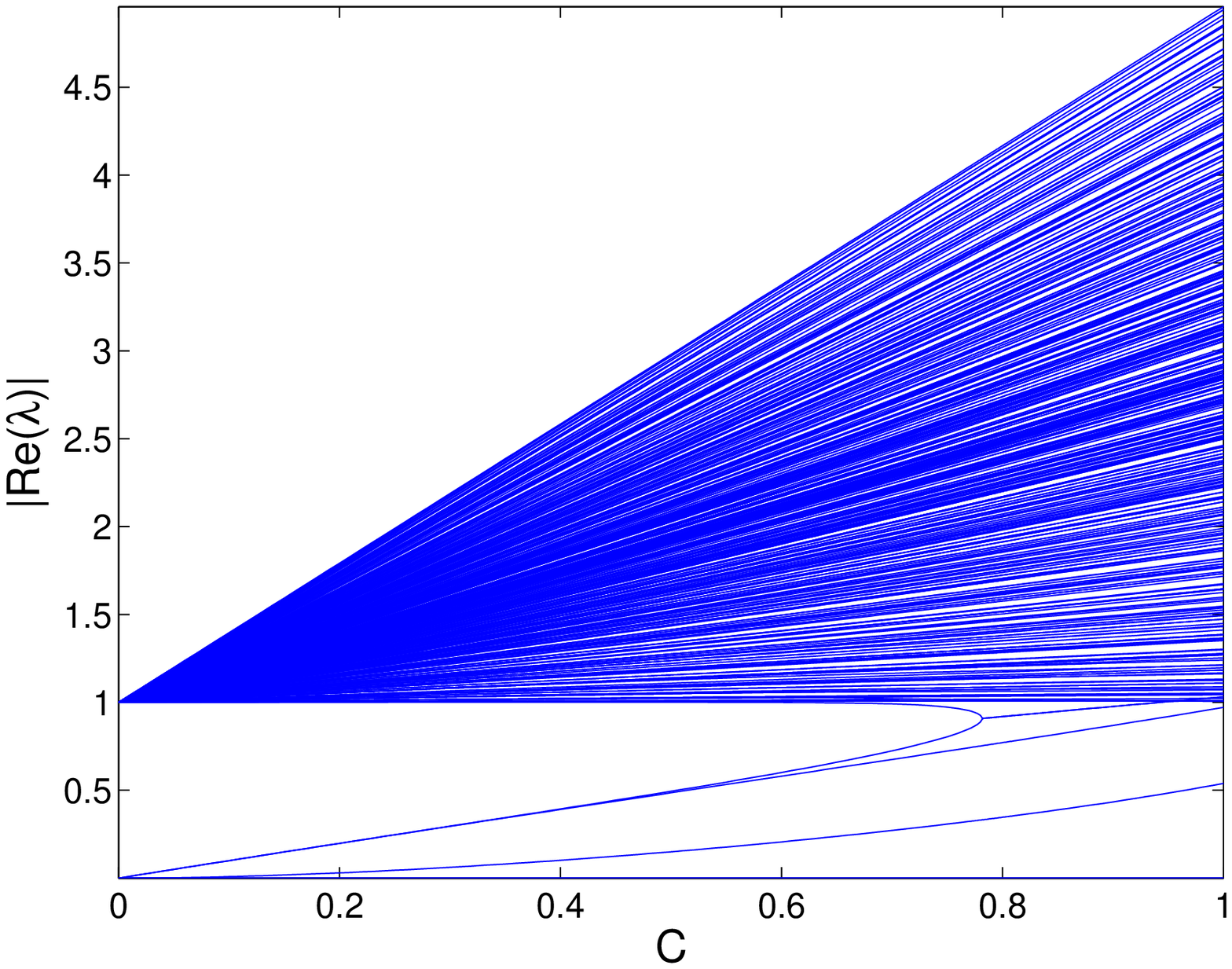} & \includegraphics[width=%
\doublefig]{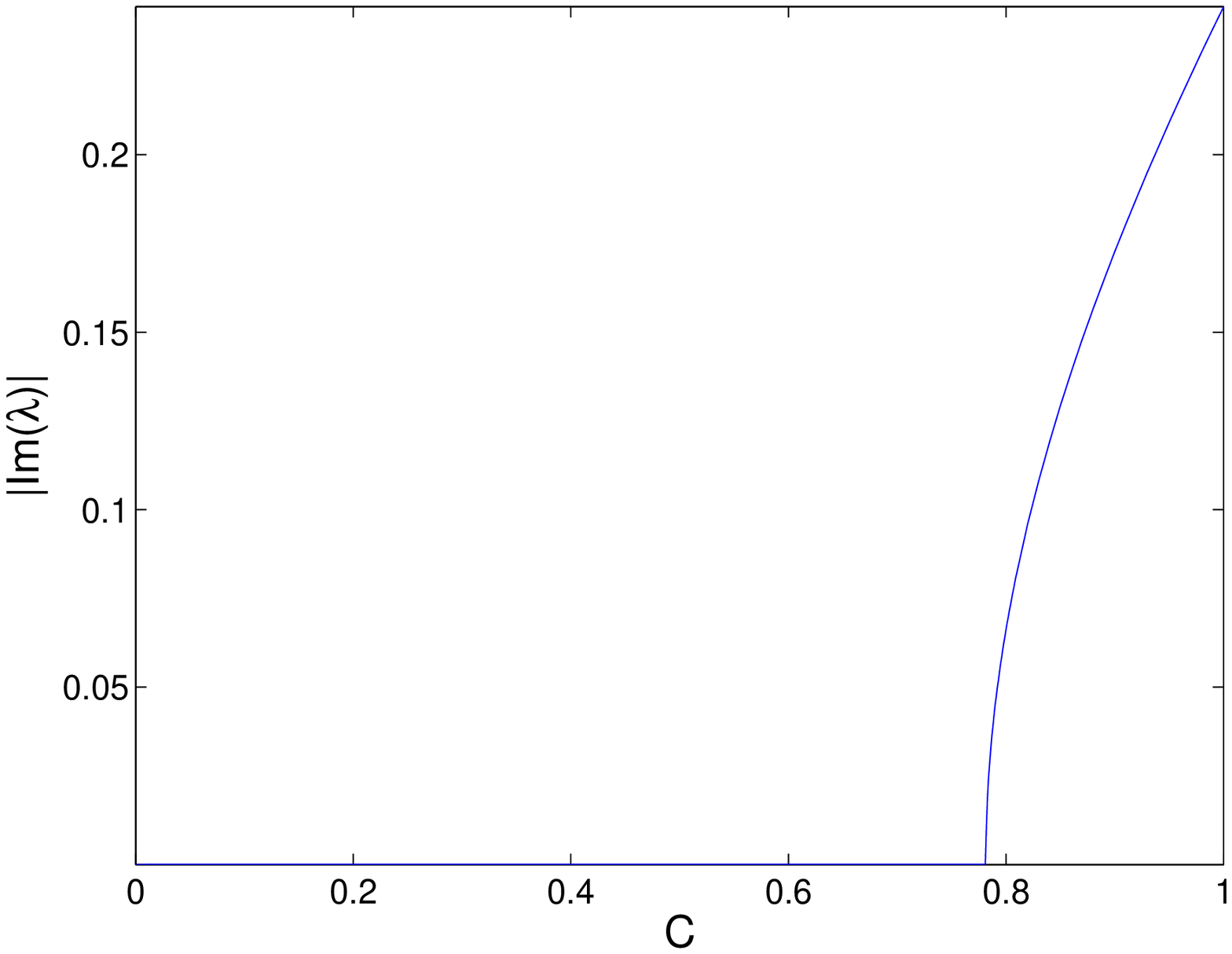} \\
\includegraphics[width=\doublefig]{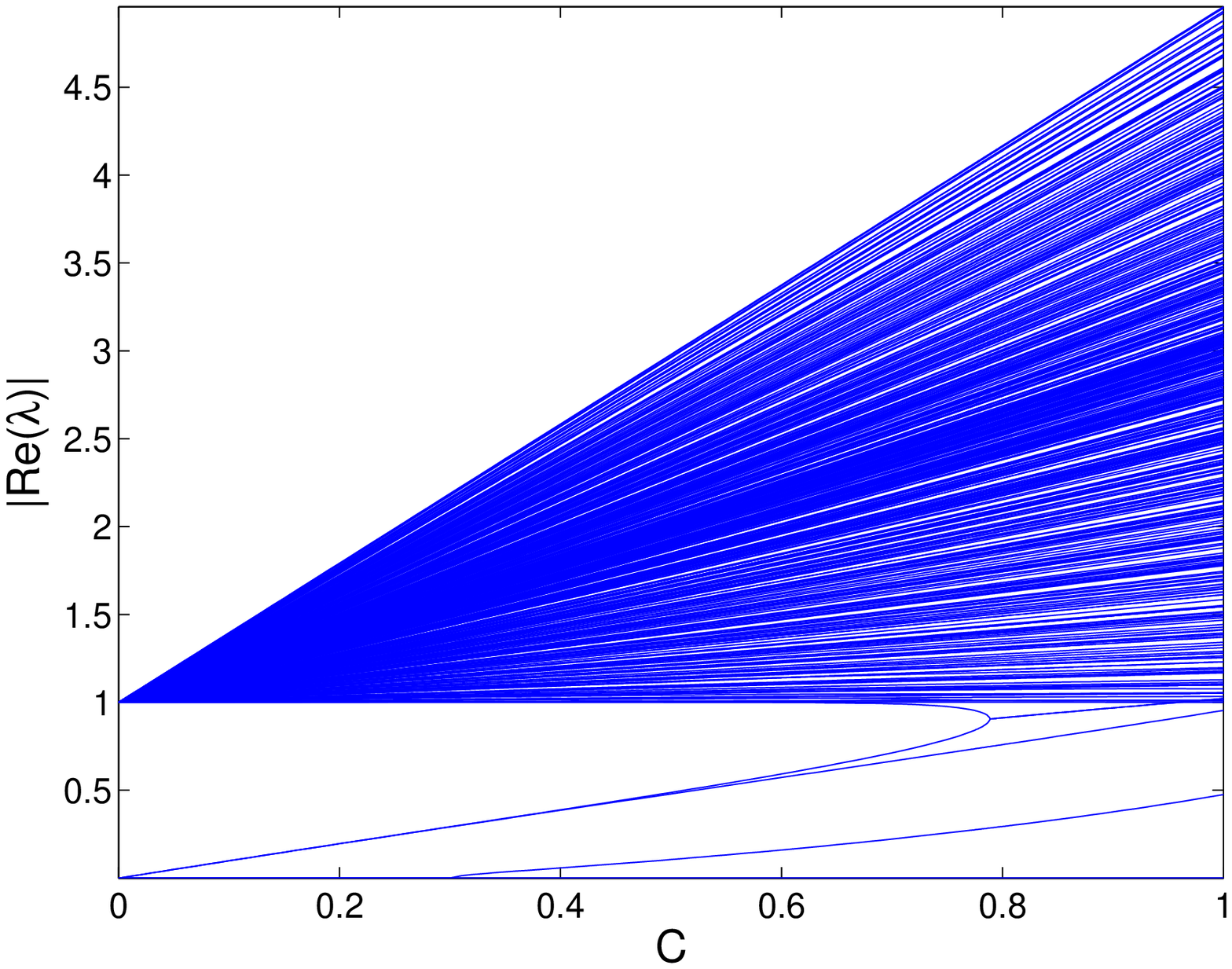} & \includegraphics[width=%
\doublefig]{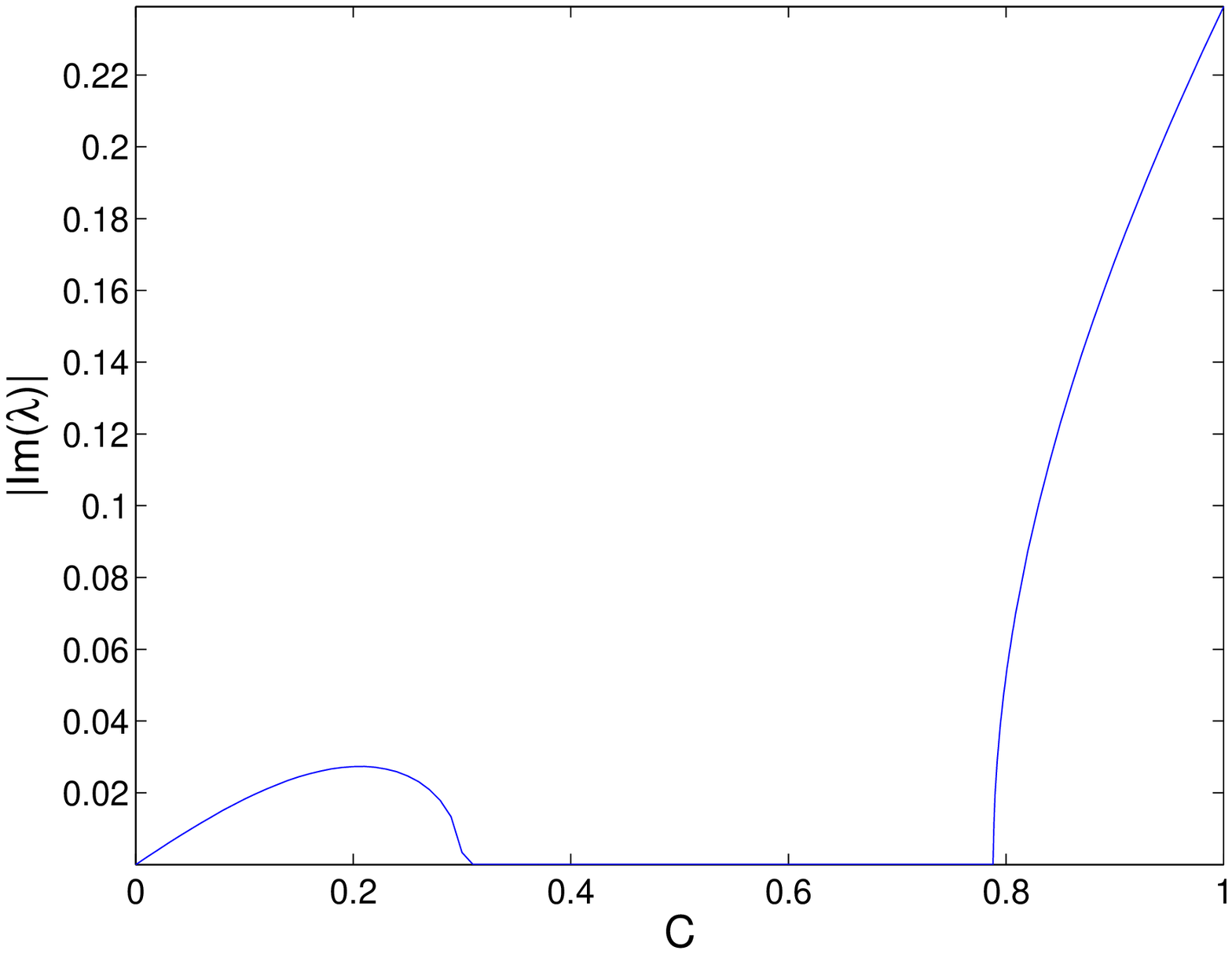} \\
\includegraphics[width=\doublefig]{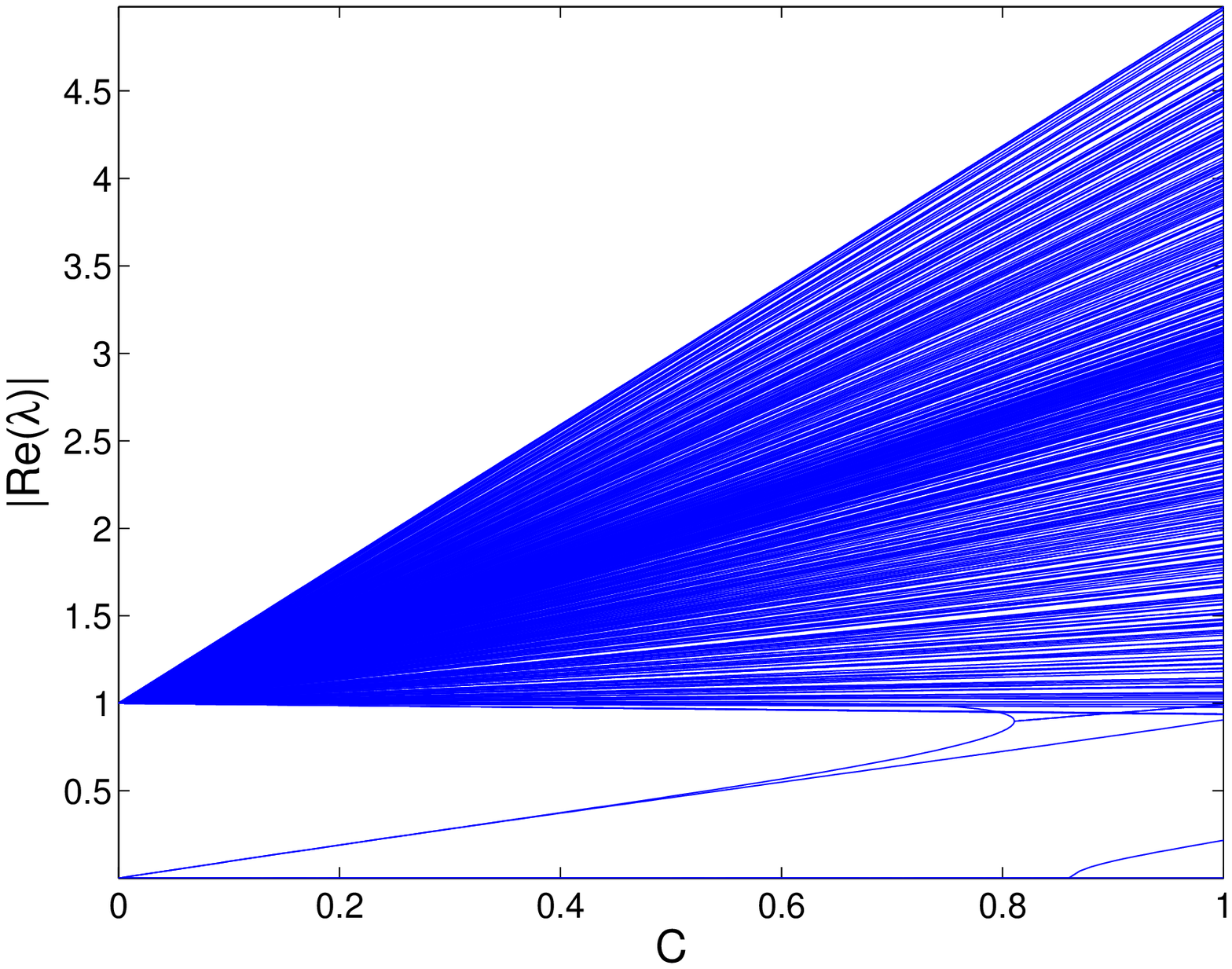} & \includegraphics[width=%
\doublefig]{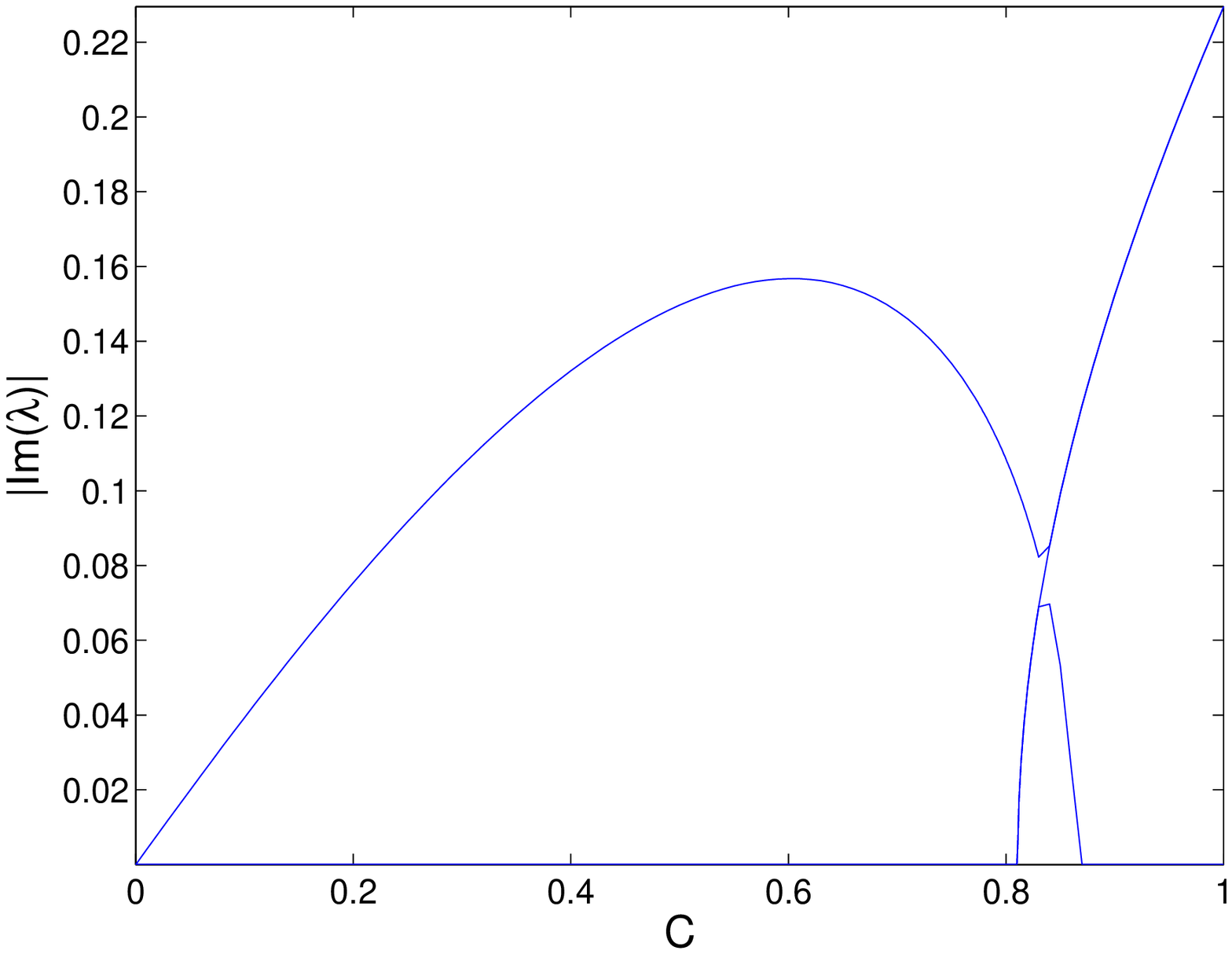}%
\end{tabular}%
\end{center}
\caption{(Color online) Dependences on $C$ of the real and imaginary parts
(left and right panels, respectively) of eigenfrequencies of small
perturbations about vortex solitons with $S=1$ in the ordinary
(non-rotating) DNLS lattice (top panels), and in the rotating one, with $%
\Omega =0.01$ (central panels) and $\Omega =0.04$ (bottom panels). In the
rotating case, the center of the vortex coincides with the rotation pivot of
the lattice.}
\label{fig:stabvort1}
\end{figure}

At a given value of $\Omega $, VSs are exponentially unstable for $C<\tilde{C%
}_{\mathrm{cr}}^{(S=1)}\,$, and they feature an oscillatory
instability, accounted for by a Hopf bifurcation, at
$C>{C}_{\mathrm{cr}}^{(S=1)}$. This situation takes place up at
$\Omega <\Omega _{\mathrm{cr}}^{(S=1)}=0.037 $. An example is given
by Fig. \ref{fig:stabvort1}, which displays the instability growth
rate of the VS, $\left\vert \mathrm{Im}(\lambda )\right\vert $,
versus $C$ for fixed $\Omega =0.01$. The plot also shows the real
part of the eigenfrequencies, $\left\vert \mathrm{Re}(\lambda
)\right\vert $, and includes, for the sake of comparison, the same
dependences in the usual DNLS equation, with $\Omega =0$. The figure
shows
that, in this case, $\tilde{C}_{\mathrm{cr}}^{(S=1)}=0.31$ and $C_{\mathrm{cr%
}}^{(S=1)}=0.788$.

\begin{figure}[tbp]
\begin{center}
\begin{tabular}{cc}
\includegraphics[width=\doublefig]{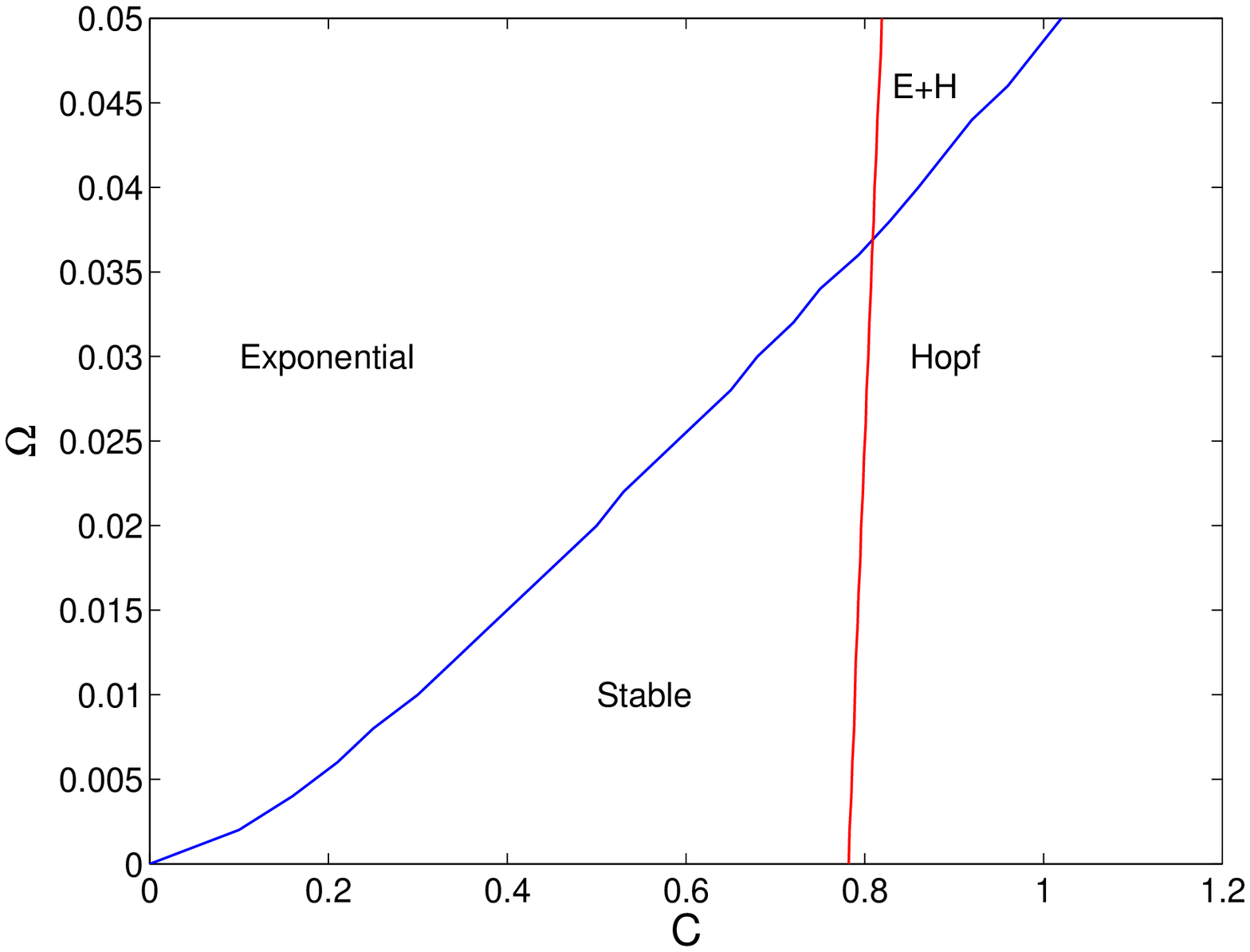} &
\end{tabular}%
\end{center}
\caption{The full stability diagram for vortex solitons with $S=1$, centered
at the rotation pivot. The E+H label holds for coexisting Hopf and
exponential instabilities.}
\label{diagram1}
\end{figure}

The overall stability region for the VSs with $S=1$ in the plane $\left(
C,\Omega \right) $ is presented in Fig. \ref{diagram1}. It is observed that $%
C_{\mathrm{cr}}^{(S=1)}$ slightly increases with $\Omega $, while the growth
of $\tilde{C}_{\mathrm{cr}}^{(S=1)}$ with $\Omega $ is fast. As a result, at
$\Omega >\Omega _{\mathrm{cr}}^{(S=1)}$ the stability region does not exist.
In the absence of the stable VSs, unstable ones feature coexistence of
exponential and Hopf instabilities. Figure \ref{fig:stabvort1} displays an
example of the latter regime, and shows that, for $\Omega =0.04$, $\tilde{C}%
_{\mathrm{cr}}^{(S=1)}=0.87$ and $C_{\mathrm{cr}}^{(S=1)}=0.81$.

In fact, the effect of the \textit{destabilization} of the VSs with $S=1$ at
small $C$ [$C<\tilde{C}_{\mathrm{cr}}^{(S=1)}$, as said above], is a
higher-order phenomenon, in terms of the expansion of the stability
eigenvalues in powers of $C$, when it is assumed small. Indeed, comparing it
with the stability analysis for the cross vortices in the ordinary model,
with $\Omega =0$ \cite{pelin2}, we observe the following. At $\Omega =0$,
modes of small perturbations around the vortex of the cross/rhombus type
feature a pair of real eigenfrequencies, with $\lambda =\pm C$ at the first
order in small $C$ (in the present notation). The same pair appears in the
present context, see Fig. \ref{fig:stabvort1}. At larger $C$, these
eigenfrequencies will give rise to the Hamiltonian-Hopf instability, upon
their collision with eigenfrequencies bifurcating from the phonon
band of linear excitations. As shown in Ref. \cite{pelin2},
the DNLS\ equation with $\Omega =0$ also gives rise to a pair of
higher-order eigenfrequencies, $\lambda \approx \pm C^{2}$ in the present
notation. The effect of the rotation forces the latter
eigenfrequencies to
separate along the imaginary axis, thus inducing the instability at small $C$%
. However, for larger $C$, the pair again becomes real, stabilizing the
configuration.
\begin{figure}[tbp]
\begin{center}
\begin{tabular}{cc}
\includegraphics[width=\doublefig]{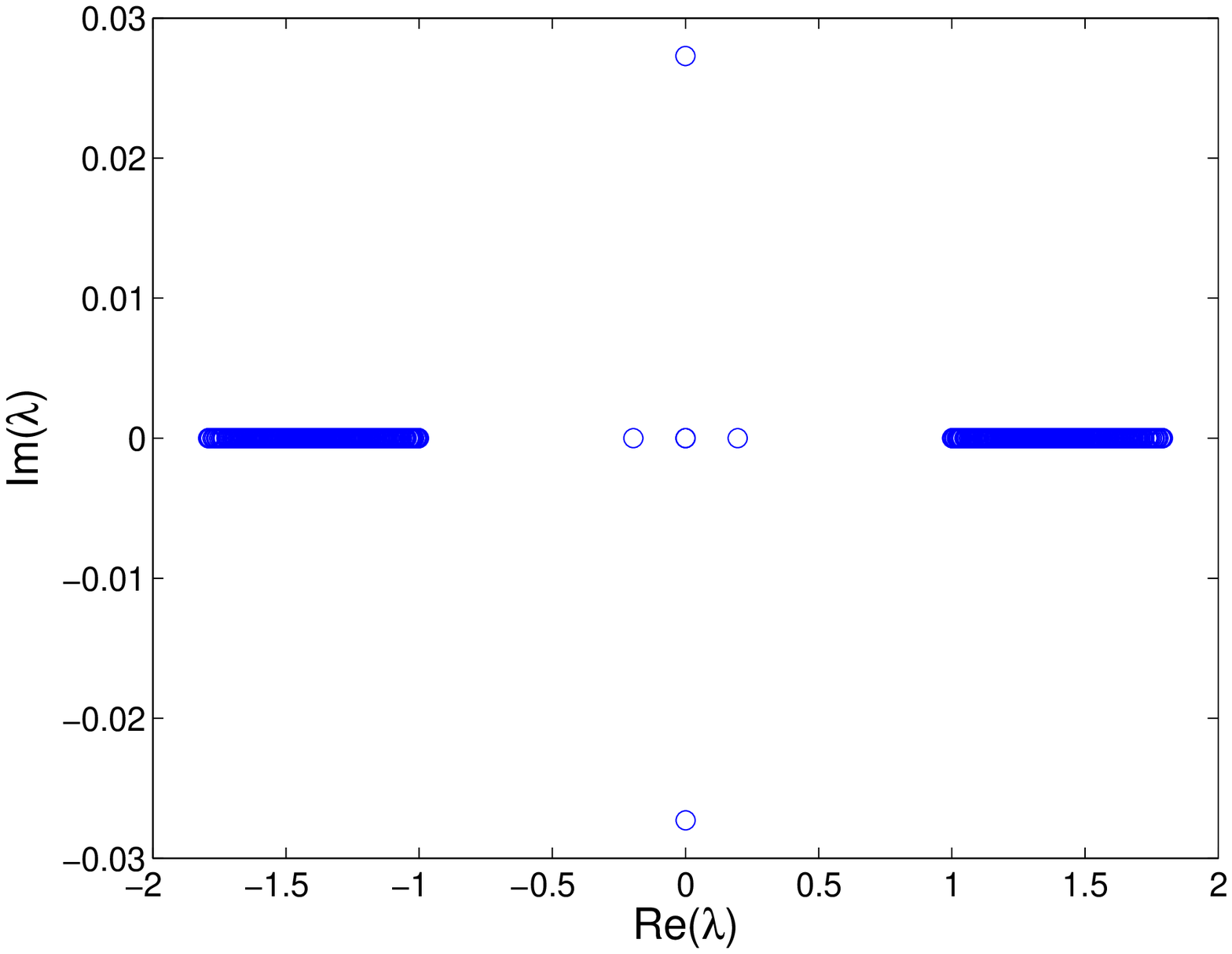} & \includegraphics[width=%
\doublefig]{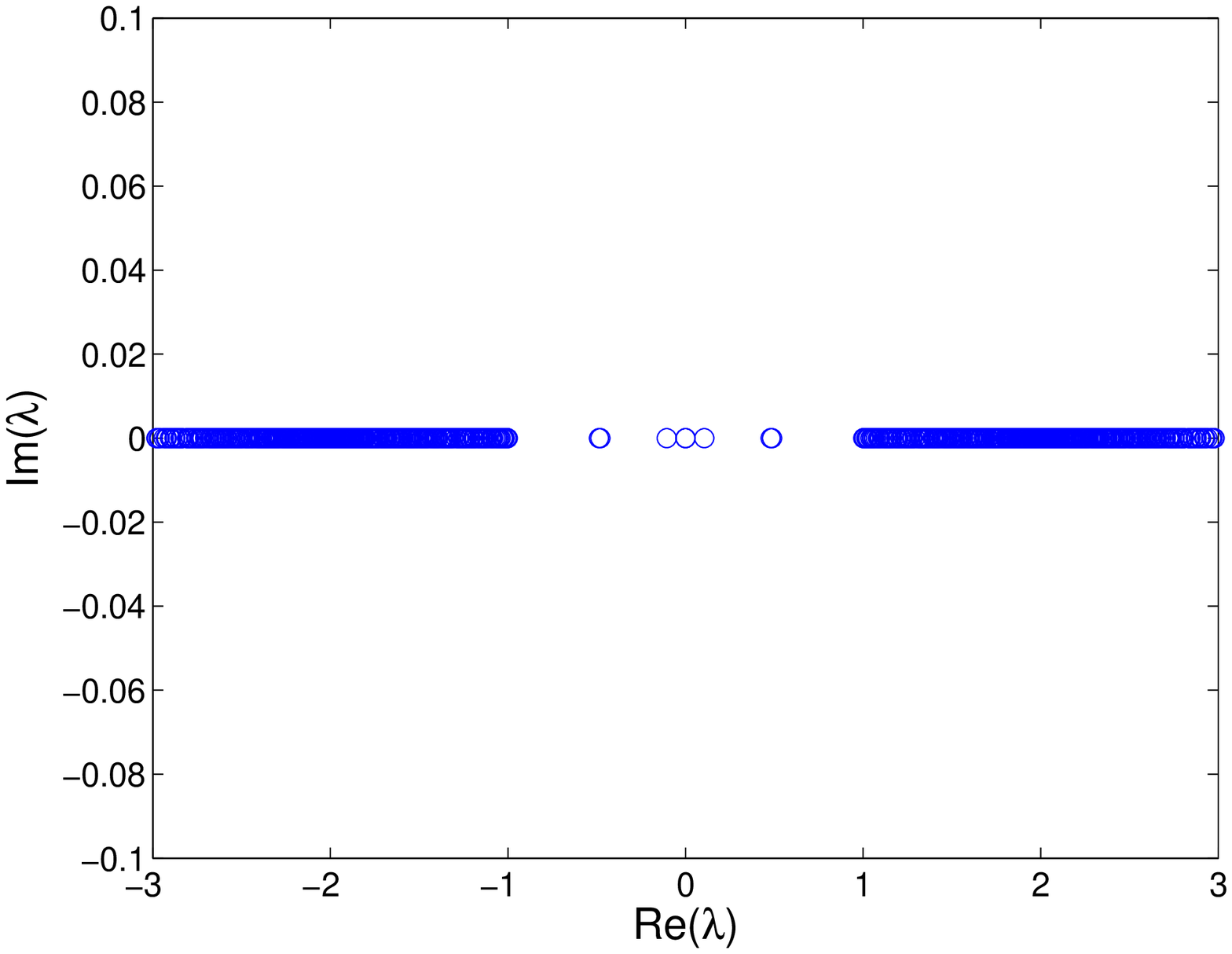} \\
\includegraphics[width=\doublefig]{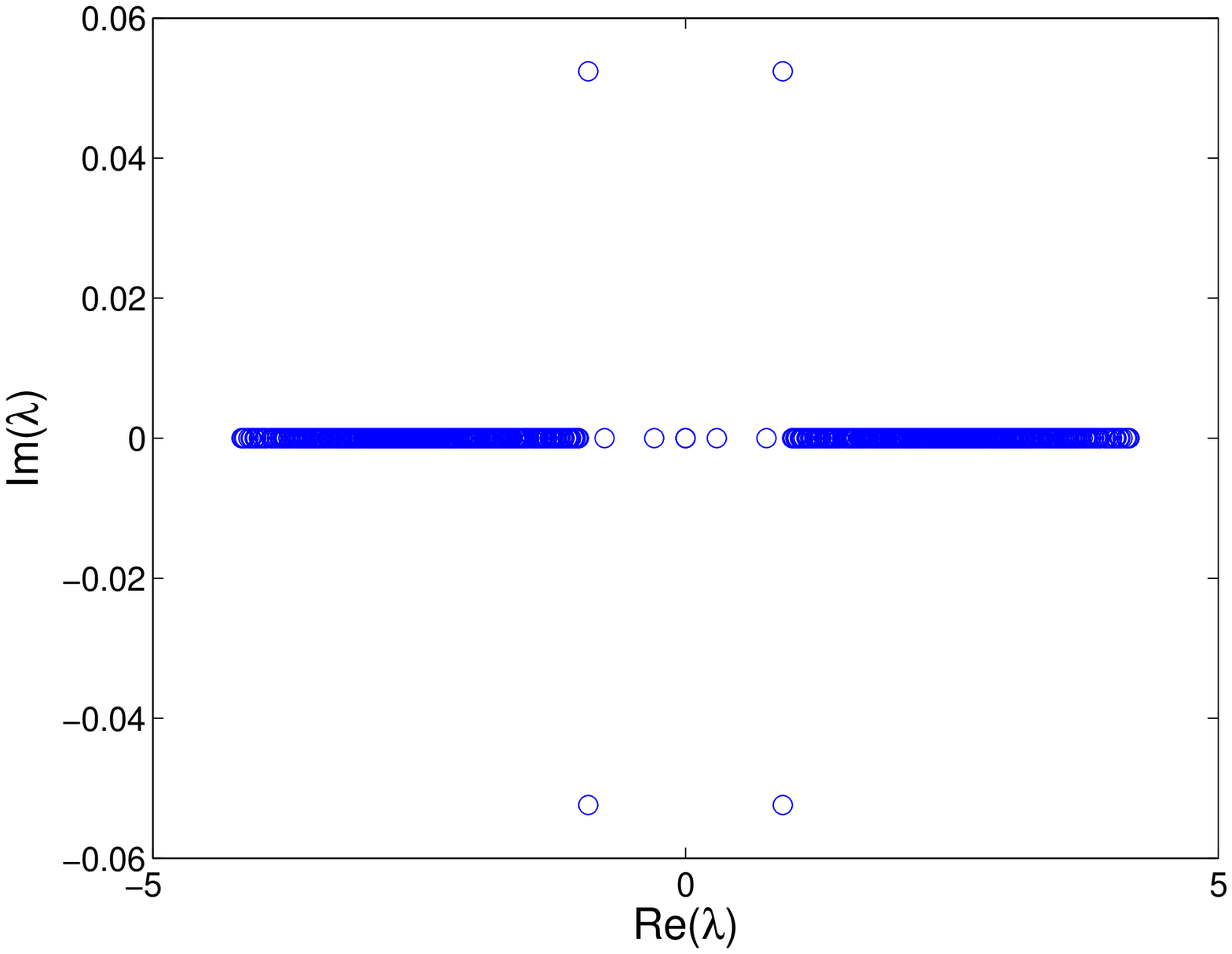} & \includegraphics[width=%
\doublefig]{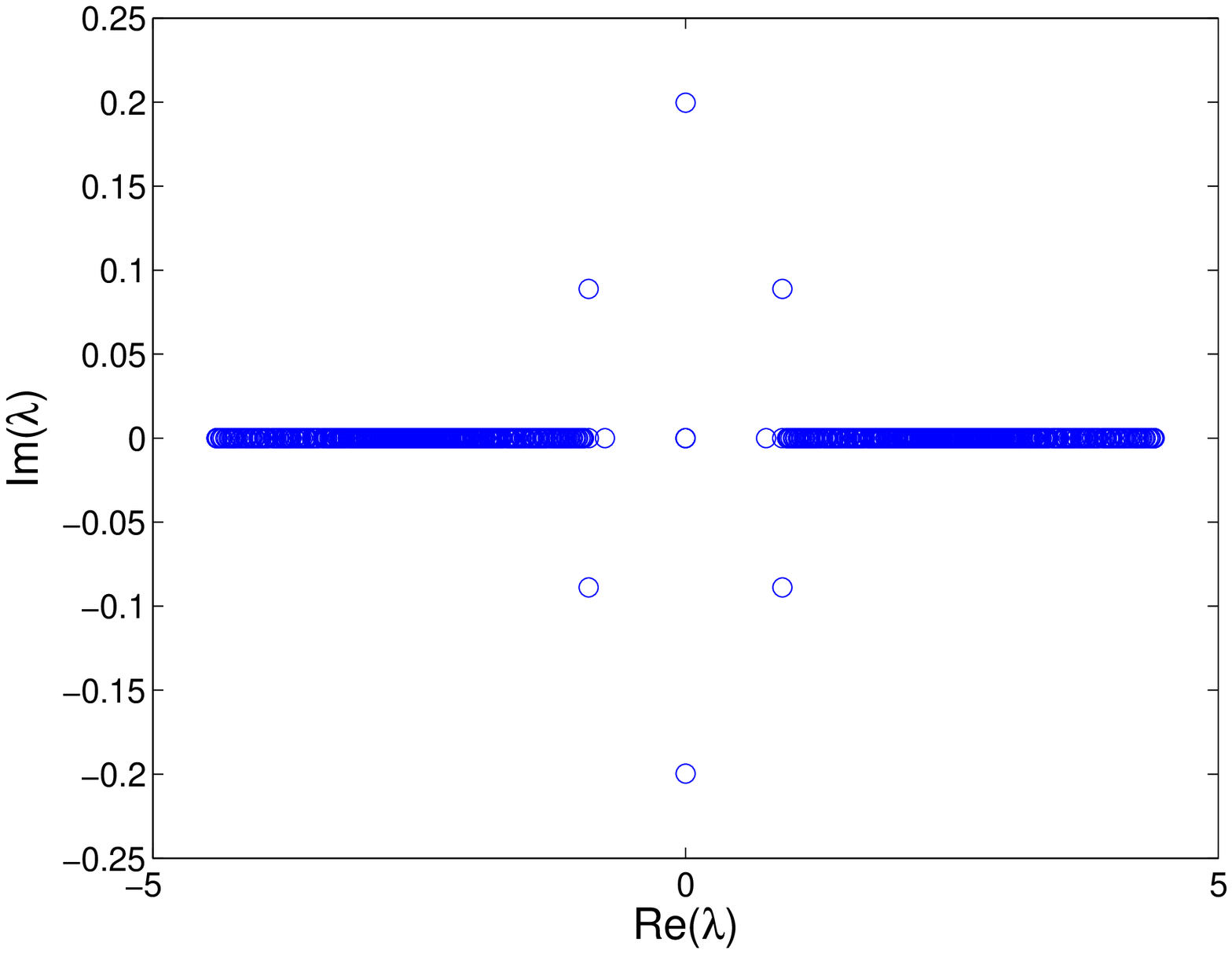}%
\end{tabular}%
\end{center}
\caption{Eigenfrequencies of small perturbations around the vortex with $S=1$
are shown for the four different cases: exponential instability
for $\Omega
=0.01$ and $C=0.2$ (top left panel); linear stability at $\Omega =0.01$, and
$C=0.5$ (top right panel); oscillatory instability at $\Omega =0.01$ and $%
C=0.8$ (bottom left panel); exponential and oscillatory instability at $%
\Omega =0.05$ and $C=0.85$ (bottom right panel).}
\label{fig:vort12}
\end{figure}

Four different regimes identified in the stability diagram in Fig. \ref%
{diagram1} are illustrated by typical examples of the stability
and instability of the VSs with $S=1$ in Fig. \ref{fig:vort12} as
follows: the exponential instability, caused by the higher-order
eigenfrequencies, as described above, is shown in the top left panel;
the top right panel shows a linearly stable case. The oscillatory
instability is presented in the bottom left panel (the latter case
is shown for sufficiently large $C$, to allow the eigenfrequencies,
which originally linearly depend on $C$, as indicated above, to collide
with eigenfrequencies bifurcating from
the phonon band). Finally, the mixed oscillatory-exponential
instability, which is possible at $\Omega >\Omega
_{\mathrm{cr}}^{(S=1)}$, is displayed in the bottom right panel.

\begin{figure}[tbp]
\begin{center}
\includegraphics[width=\doublefig]{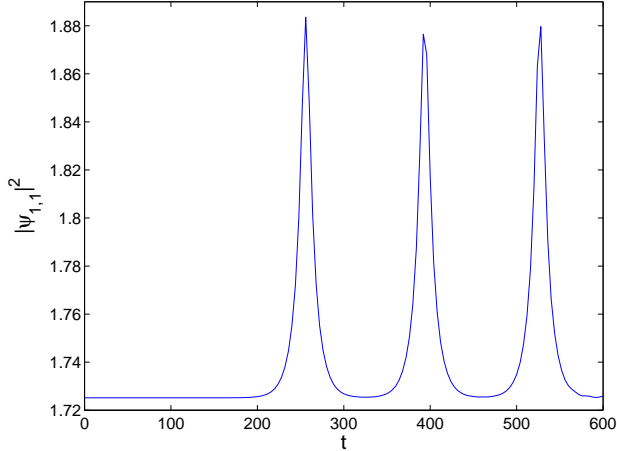}
\end{center}
\caption{The time dependence of the amplitude of a perturbed unstable vortex
soliton with $S=1$, $\Omega =0.03$, and $C=0.4$. Note that the point
corresponding to this soliton is located in the left instability region in
Fig. \protect\ref{diagram1}, i.e., the vortex is unstable to non-oscillatory
perturbations (see also the text).}
\label{fig:pulsonvort1}
\end{figure}

Since the destabilization of the VS at $C>C_{\mathrm{cr}}^{(S=1)}$ is
accounted for by the Hopf bifurcation, as confirmed by Fig. \ref%
{fig:stabvort1}, this unstable VS is transformed into a persistent\textit{\ }%
breather (not shown here), which loses the vortical structure, i.e., is
similar to the FS \cite{vort1}.
On the other hand, the destabilization at $C<%
\tilde{C}_{\mathrm{cr}}^{(S=1)}$ occurs, as seen in Figs. \ref{fig:stabvort1}%
, via a pair of imaginary eigenfrequencies with zero real parts,
i.e., an exponential instability. Its nonlinear development
eventually leads to a persistently pulsating localized state with
zero vorticity, as illustrated by Fig. \ref{fig:pulsonvort1}. The
transformation to a FS state is also observed in the region of the
coexistence of exponential and oscillatory instabilities.

\subsection{Vortex solitons with $S=2$}

\begin{figure}[tbp]
\begin{center}
\begin{tabular}{cc}
\includegraphics[width=\doublefig]{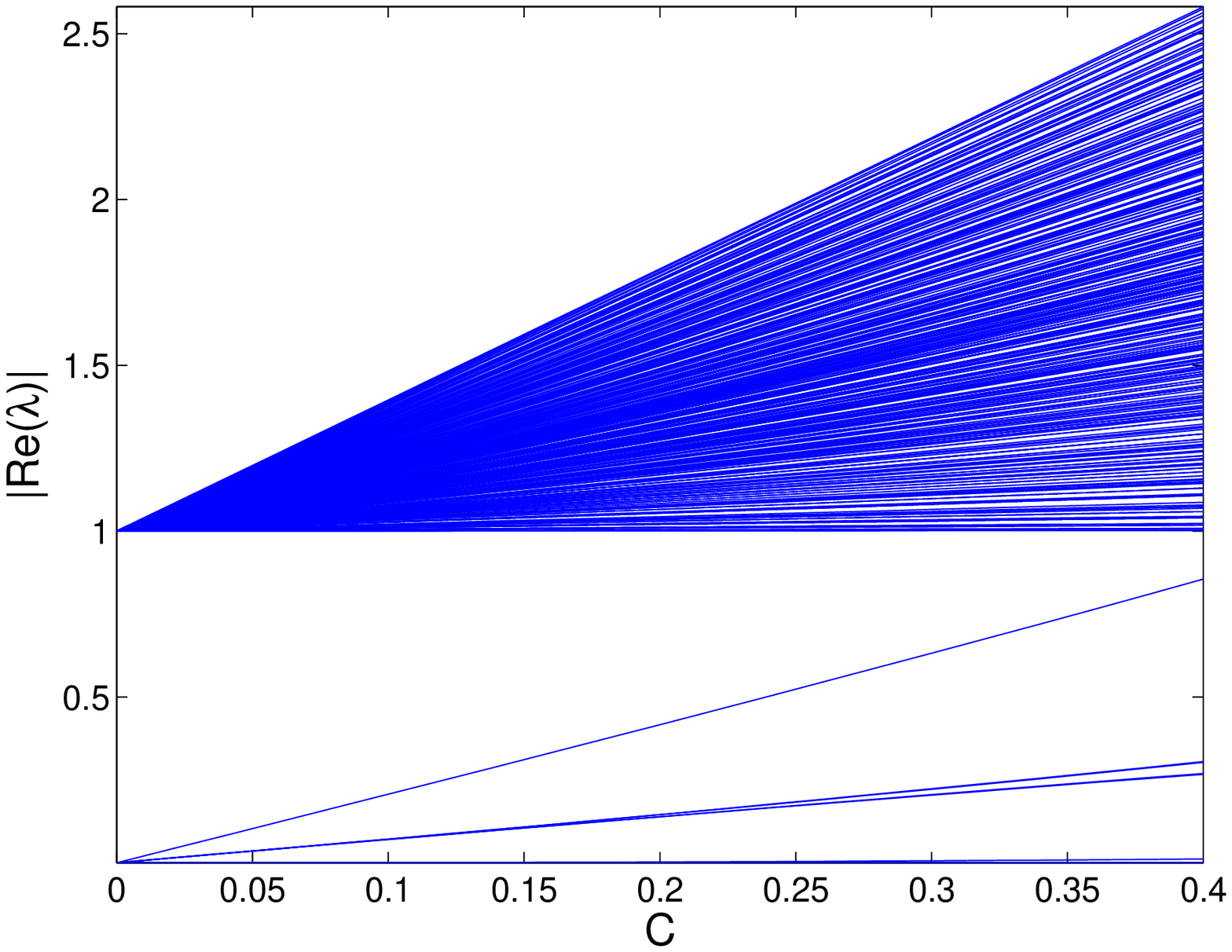} & \includegraphics[width=%
\doublefig]{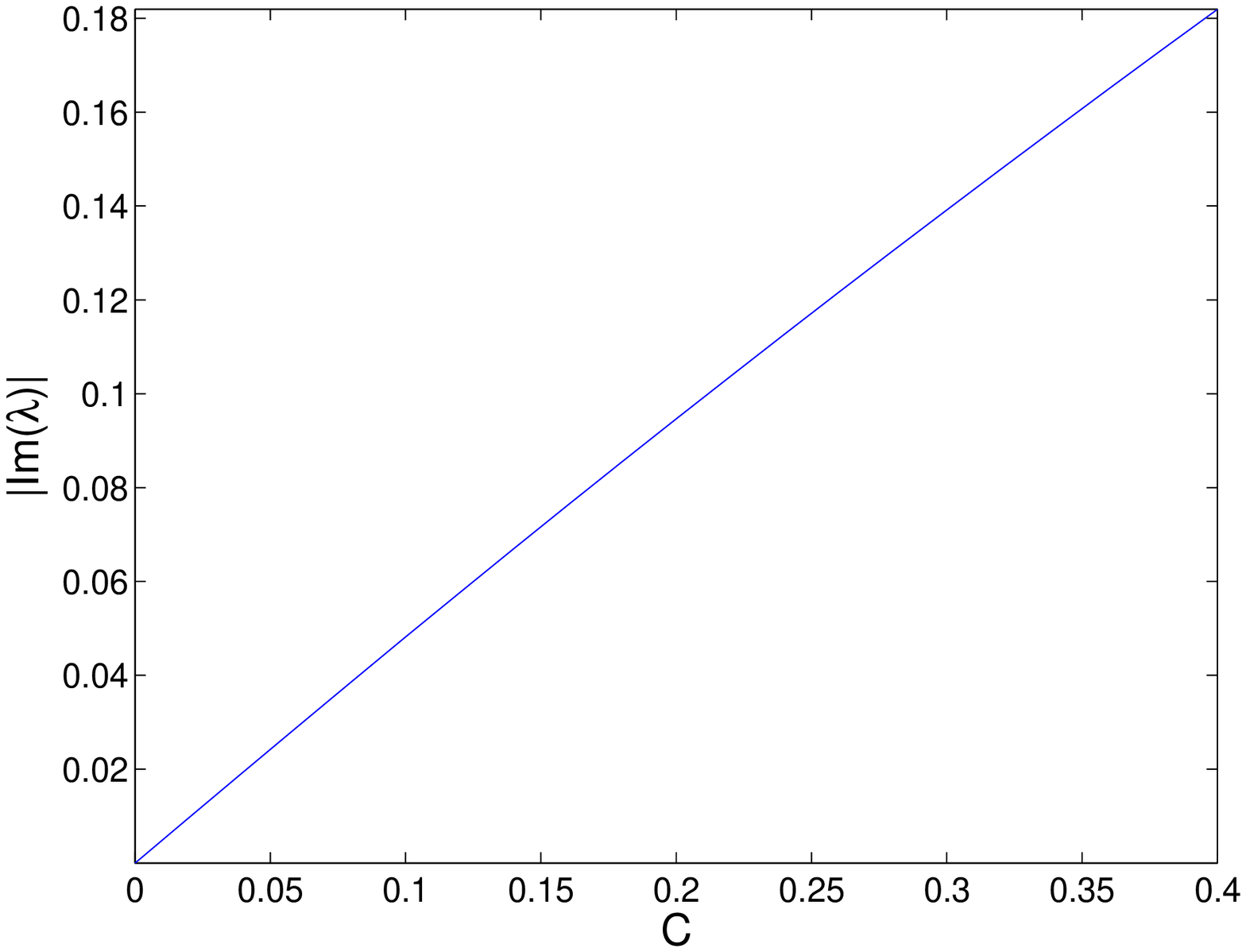} \\
\includegraphics[width=\doublefig]{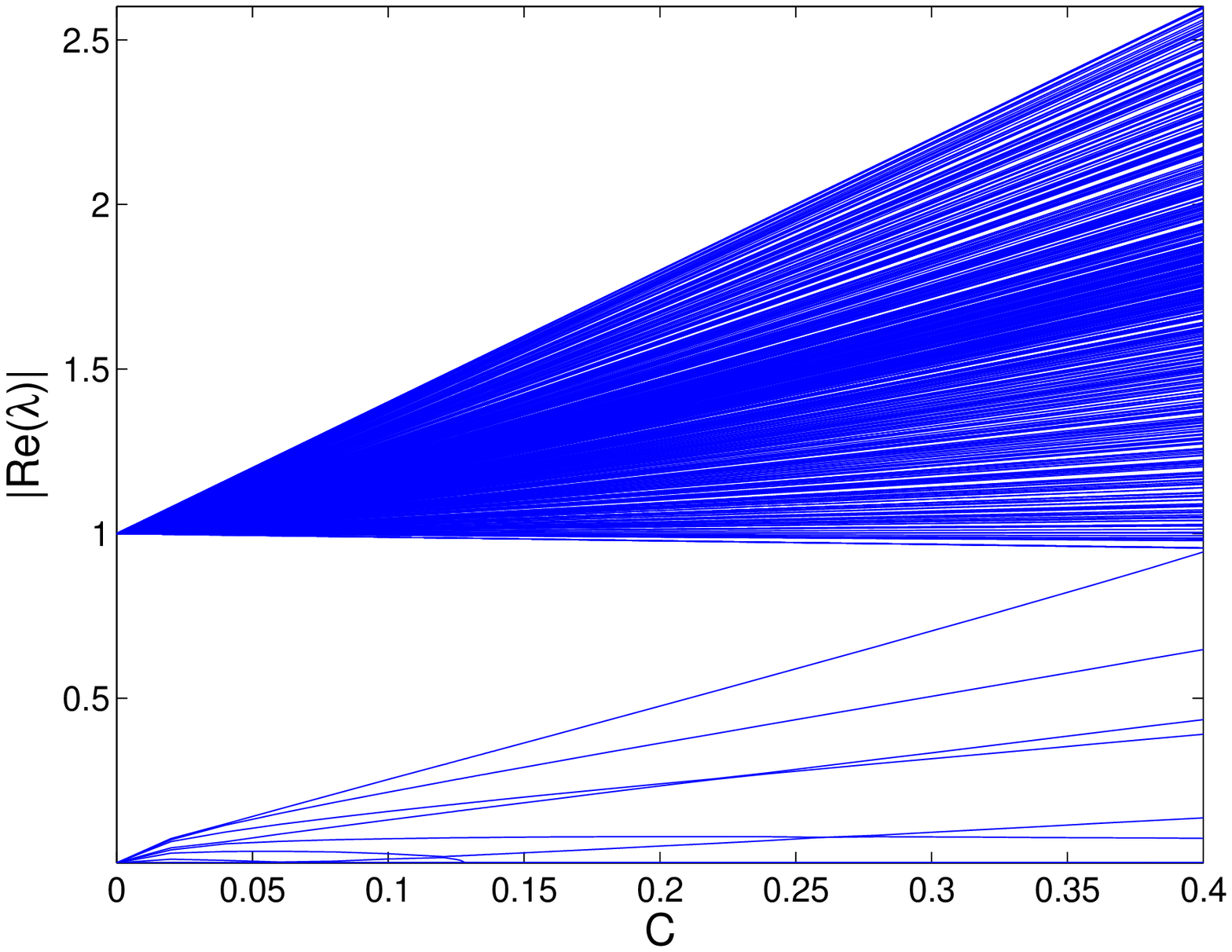} & \includegraphics[width=%
\doublefig]{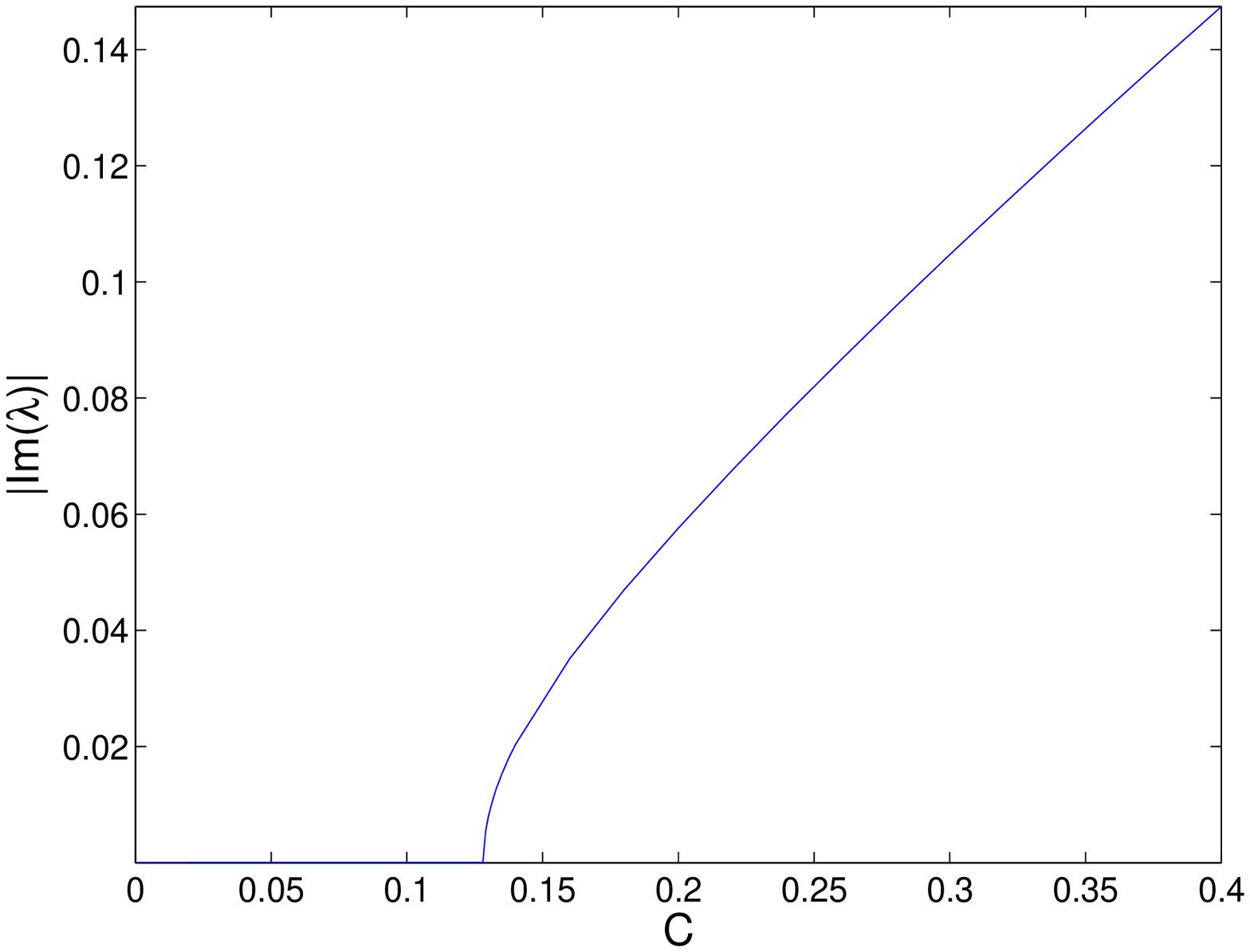}%
\end{tabular}%
\end{center}
\caption{(color online) The same as in Fig. \protect\ref{fig:stabvort1}, but
for the vortex solitons with $S=2$. Top and bottom panels correspond to $%
\Omega =0$ and $\Omega =0.05$, respectively. Notice the contrast between the
absence of instabilities at small $C$ in the bottom right panel and the
immediate destabilization in the top right panel.}
\label{fig:stabvort2}
\end{figure}

Another way in which the rotating lattice drastically alters the stability
features of the ordinary 2D DNLS model concerns the VSs with $S=2$. At $%
\Omega =0$, all such localized vortices are unstable due to an
imaginary eigenfrequency proportional (at the leading order) to
coupling constant $C$ \cite{vort1,vort2}, see the top panel of
Fig. \ref{fig:stabvort2} (which may be compared with Fig. 4 of
\cite{vort2}). In the rotating lattice, the solitons with $S=2$
acquire a finite \emph{stability region}, as manifested by the
example displayed in Fig. \ref{fig:stabvort2}. Note that, unlike
the stability interval for the VSs with $S=1$, see Eq.
(\ref{VSstability}), only an upper stability border exists for the
$S=2$ solitons, i.e., the respective stability interval is
$0<C<C_{\mathrm{cr}}^{(S=2)}$. For instance, in the case shown in
Fig. \ref{fig:stabvort2}, the stability border induced by the
rotation is $C_{\mathrm{cr}}^{(S=2)}=0.12$. Note that the
mechanism of the stabilization of the $S=2$ VSs in the present
model is different from that reported in Ref. \cite{s2vort}, where
localized vortices with $S=2$ were stabilized by an impurity
(\textit{inert site}) placed at the center. In that case, the
unstable eigenmode was suppressed by the defect, making all
eigenfrequencies real; eventual destabilization occurred due to
collisions of those real eigenfrequencies with the linear spectrum.
Here, the
rotation affects the unstable (imaginary) eigenfrequency of the $S=2$ VS,
rendering it real for small $C$. However, as $C$ is increased
the eigenfrequency eventually becomes imaginary again, leading
to the instability of the VS.


\begin{figure}[tbp]
\begin{center}
\includegraphics[width=\doublefig]{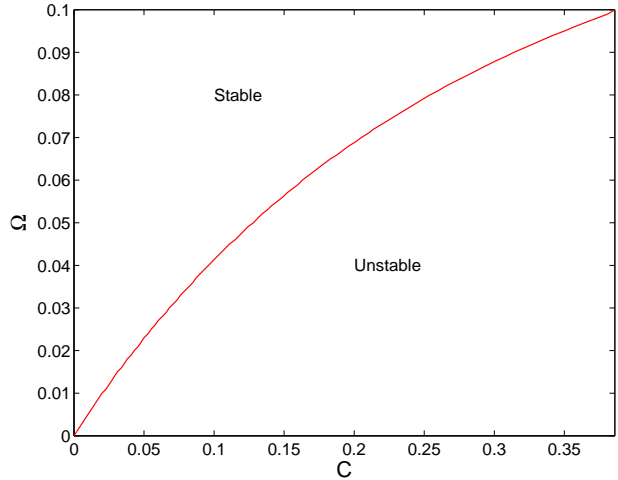}
\end{center}
\caption{The stability diagram, similarly to Fig. \protect\ref{diagram1},
but for vortex solitons with $S=2$.}
\label{diagram2}
\end{figure}

The stability diagram for the VSs with $S=2$ in the $\left( C,\Omega \right)
$ plane is presented in Fig. \ref{diagram2}, indicating the increasing
stabilization effect of larger rotation frequencies. As in the case of $S=1$%
, the instabilities of the VSs with $S=2$ transform it into a persistent
breather, but without the vortical structure. In fact, a similar qualitative
conclusion was made in the ordinary model, with $\Omega =0$, where all VSs
with $S=2$ are unstable \cite{vort1,vort2}.

\subsection{Quadrupole and octupole solitons}

Along with complex solutions for localized vortices with $S\geq 2$, the
ordinary 2D DNLS equation, with $\Omega =0$, gives rise to real solutions in
the form of quadrupoles and octupoles, that resemble higher-order vortices,
but carry no topological charge \cite{higher-order-vortex,supervortex}. In
particular, quadrupoles, which include four lattice sites with alternating
phases, have their stability region, $0<C\leq C_{\mathrm{cr}}^{\mathrm{(quad)%
}}$, in the model with $\Omega =0$; the same is true for octupoles, which
are based on eight sites with alternating phases. In the anti-continuum
limit, the quadrupole and octupole solutions are seeded, respectively, by
the following configurations:, $\phi _{0,\pm 1}=1,\phi _{\pm 1,0}=-1$, and $%
\phi _{0,2}=\phi _{1,-1}=\phi _{2,1}=\phi _{-1,0}=1,\phi _{1,2}=\phi
_{0,-1}=\phi _{-1,1}=\phi _{2,0}=-1$, all other sites having $\phi _{m,n}=0$%
. Note that the latter configuration is tantamount to the necklace soliton
pattern, that was recently observed in a photorefractive crystal with a
photoinduced lattice \cite{neck}.

We have constructed families of quadrupole and octupole soliton solutions in
the present model with $\Omega >0$, proceeding from the anti-continuum
patterns to finite $C$. When doing so, the rotation pivot in Eq. (\ref{phi})
was set at the center of the respective pattern, i.e., we set $\xi =\upsilon
=0$ and $\xi =\upsilon =1/2$, for the quadrupole and octupole solutions
respectively. It was found that the critical values of the coupling
constant, $C_{\mathrm{cr}}^{\mathrm{(quad)}}$ and $C_{\mathrm{cr}}^{\mathrm{%
(oct)}}$, which border the stability regions for both these families, very
weakly depend on $\Omega $, unlike the situation for the VSs with $S=2$, cf.
Fig. \ref{diagram2}, but quite similar to what was found above for the
localized vortices with $S=1$, see the right stability border in Fig. \ref%
{diagram1}.

\section{Conclusion}

The objective of the present work was to introduce a discrete
version of the 2D model combining the self-attractive cubic
nonlinearity and rotating square-lattice potential. The discrete
model can be implemented in BEC stirred by a rotating strong optical
lattice, or, in principle, also in a twisted bundle of nonlinear
optical fibers. Localized solutions of two types were considered:
off-axis FSs (fundamental solitons), with the center placed at
distance $R$ from the rotation pivot, and on-axis VSs (vortex
solitons), with vorticity $S=1$ and $2$. For the FSs, the stability
interval was found in the form of $0<C<C_{\mathrm{cr}}(R)$, where
$C$ is the coupling constant of the discrete lattice, and
$C_{\mathrm{cr}}(R)$ a monotonically decreasing function, see Fig.
\ref{fig:Ccr}. A qualitative explanation to this result was
proposed, based on the analysis of the balance between the
lattice-pinning and centrifugal forces. For VSs with $S=1$, the
dependence of the stability region on rotation frequency $\Omega $
was found, in the form of Eq. (\ref{VSstability}) and Fig.
\ref{diagram1}, with the conclusion that the stability is only
possible for $\Omega <\Omega _{\mathrm{cr}}$. A key feature, which
makes the situation different from earlier stability analyses of
such structures in the standard 2D DNLS model, is a higher-order
eigenfrequency, shifted towards instability for sufficiently weak
lattice coupling, in the presence of rotation. On the other hand,
VSs with $S=2$, which are always unstable in the model with $\Omega
=0$, are stabilized by
the rotation in region $0<C<C_{\mathrm{cr}}^{(S=2)}$, as shown in Fig. \ref%
{diagram2}. For the vortices with $S=2$, the reverse effect happens in
comparison with $S=1$, namely an unstable (at $\Omega =0$) eigenfrequency is
tipped by the rotation in the opposite direction; i.e.,
for $S=1$ a real eigenfrequency becomes imaginary in the presence of
$\Omega$, while the reverse is true for $S=2$. Quadrupole and octupole
solitons, with the center coinciding with the pivotal point, were briefly
considered too, with a conclusion that their stability regions are almost
the same as in the ordinary model (with $\Omega =0$). An estimate for
relevant values of $\Omega $ in physical units was given for both physical
realizations of the model, i.e., the rotation frequency of the optical
lattice stirring the self-attractive BEC, and the pitch of the twisted
bundle of optical fibers.

The analysis initiated in this work can be developed in several directions.
In particular, results obtained in the continuum model considered in Ref.
\cite{HS} suggest that, at sufficiently large $\Omega $, the fully localized
FSs, with the center shifted off the axis (i.e., continuum counterparts of
the discrete FSs considered in the present work), are unstable (or do not
exist), while stable ring-shaped solitons may appear instead, with the
center of the ring coinciding with the pivot. In fact, the shape of these
zero-vorticity rings may be similar to that of the vortices considered above.

Another interesting direction would be to apply techniques elaborated in
Refs. \cite{vort2} and \cite{pelin} to this considerably more difficult
setting. It would be especially relevant to repeat the calculation of the
eigenfrequencies
for the vortices with $S=1$ and $S=2$ by means of those methods
in the presence of rotation, and quantify the impact of $\Omega $ on the
eigenvalues, to rigorously investigate some effects which were outlined
above in a qualitative form.

Another possibility is to consider a discrete limit of the model with the
rotating quasi-1D potential, such as the one with the repulsive cubic
nonlinearity, which was introduced  in Ref. \cite{Barcelona}. Studies along
these directions are currently underway and will be reported elsewhere.

\noindent \textbf{Acknowledgements}. J.C. acknowledges financial support
from the MECD project FIS2004-01183. The work of B.A.M. was partially
supported by the Israel Science Foundation through Excellence-Center Grant
No. 8006/03, and by German-Israel Foundation through Grant No.
I-884-149.7/2005. PGK gratefully acknowledges the support of
NSF-DMS-0505663, NSF-DMS-0619492 and NSF-CAREER.

\end{document}